\documentclass[fleqn,usenatbib]{mnras}

\usepackage[T1]{fontenc}

\DeclareRobustCommand{\VAN}[3]{#2}
\let\VANthebibliography\thebibliography
\def\thebibliography{\DeclareRobustCommand{\VAN}[3]{##3}\VANthebibliography}

\usepackage{graphicx}	
\usepackage{amsmath}	
\usepackage{amssymb}	

\newcommand {\ha} {H$\alpha$}

\newcommand{\hi}{H\,\textsc{i}}

\newcommand {\kpc} {\,{\rm kpc}}
\newcommand {\Mpc} {\,{\rm Mpc}}

\newcommand {\kmsMpc} {\,{\rm km\,s}^{-1}\,{\rm \Mpc}^{-1}}

\newcommand {\de}{^{\circ}}

\newcommand {\msun}{\,{\rm M}_\odot}

\newcommand{\Myr}{\,{\rm Myr}}
\newcommand{\Gyr}{\,{\rm Gyr}}

\newcommand{\avg}[1]{\left< #1 \right>} 
\def\aj{AJ}%
\def\apj{ApJ}%
\def\apjl{ApJ}%
\def\apjs{ApJS}%
\def\ao{Appl.~Opt.}%
\def\aap{A\&A}%
\def\aapr{A\&A~Rev.}%
\def\aaps{A\&AS}%
\def\mnras{MNRAS}%
\def\pasp{PASP}%
\def\nat{Nature}%
\usepackage{newtxtext,newtxmath}

\title[Ageing of quenched spirals]{The morphological transformation of ram pressure stripped galaxies: a pathway from late to early galaxy types}

\author[A. Marasco et al.]{
A. Marasco,$^{1}$\thanks{E-mail: antonino.marasco@inaf.it}
B.\,M. Poggianti,$^{1}$
J. Fritz,$^{2}$
A. Werle,$^{1}$
B. Vulcani,$^{1}$
A. Moretti,$^{1}$
M. Gullieuszik,$^{1}$
A. Kulier,$^{1}$
\\
$^{1}$INAF - Padova Astronomical Observatory, Vicolo dell’Osservatorio 5, 35122 Padova, Italy\\
$^{2}$ Instituto de Radioastronomia y Astrofisica, UNAM, Campus Morelia, AP 3-72, CP 58089, Mexico\\
}

\date{Accepted XXX. Received YYY; in original form ZZZ}

\pubyear{2023}

\begin{document}
\label{firstpage}
\pagerange{\pageref{firstpage}--\pageref{lastpage}}
\maketitle

\begin{abstract}
We investigate how the ageing of stellar populations can drive a morphological transformation in galaxies whose star formation (SF) activity has been quenched on short timescales, like in cluster galaxies subject to ram pressure stripping from the intracluster medium.
For this purpose, we use a sample of 91 galaxies with MUSE data from the GASP program and of their spatially resolved SF history derived with the spectral modelling software \textsc{sinopsis}.
We simulate the future continuation of the SF activities by exploring two quenching scenarios: an instantaneous truncation of the SF across the whole disc, and an outside-in quenching with typical stripping timescales of $0.5\Gyr$ and $1\Gyr$.
For each scenario we produce mock MUSE spectroscopic datacubes and optical images for our galaxies during their evolution, and classify their morphology using a new diagnostic tool, calibrated on cluster galaxies from the OmegaWINGS Survey.
We find that, in all scenarios considered, the initial galaxy population dominated by blue-cloud spirals ($\sim90\%$) evolves into a mixed population mostly composed by red-sequence spirals ($50-55\%$) and lenticulars ($\sim40\%$).
The morphology transformation is completed after just $1.5-3.5$ Gyr, proceeding faster in more efficient quenching scenarios.
Our results indicate that, even without accounting for dynamical processes, SF quenching caused by the harsh environment of a cluster can significantly affect the morphology of the infalling galaxy population on timescales of a few Gyr.
\end{abstract}




\section{Introduction}\label{s:intro}
Galaxy morphology is thought to be intimately connected to galaxy mass assembly history, with disc-like structures largely arising from the slow cooling of high angular momentum gas onto dark matter halos \citep[e.g.][]{FallEfstathiou80,Mo+98}, and spheroidal systems formed either via the monolithic collapse of massive clouds with low angular momentum \citep[e.g.][]{Larson69} or via dissipationless mechanisms such as dry mergers \citep[e.g.][]{Cole+2000}.
Understanding the formation mechanisms of the different morphology types and assessing the physical processes that drive the transition from one type to another are crucial areas of interest in modern extragalactic astrophysics.

Observationally, morphological fractions are strongly influenced by two factors.
One is the galaxy stellar mass, $M_\star$.
The existence of a morphology-mass relation, where the fraction of early-type galaxies steadily grows with increasing $M_\star$ at the expense of the late-type population, is well established \citep[e.g.][]{Desai+07,Bernardi+10,Vulcani+11,Calvi+12,Kelvin+14,Moffett+16} and is a strong indication of the diverse assembly history associated with systems of different $M_\star$.
The second factor is the environment in which the galaxy resides.
The observational evidence for a prevalence of early-type galaxies in nearby clusters, contrasted with a prevalence of disc-like systems in isolation, dates back to almost a century ago \citep{HubbleHumason31}.
\citet{Oemler74} and \citet{Dressler+80} were the firsts to establish the existence of a morphology–density relation in clusters in the local Universe, indicating a steady increase in the population of elliptical and lenticular (or S0) galaxies, and a corresponding decrease of spirals, for increasing density.

The advent of the \emph{Hubble Space Telescope} (HST) made possible to study the evolution of such relations with redshift: different studies \citep{Dressler+97,Fasano+00,Treu+03,Postman+05,Desai+07, Vulcani+11, Vulcani+11b} revealed that spirals are proportionally much more common (a factor of $\sim2\!-\!3$), and lenticulars much rarer, in distant (up to $z\sim0.8$) than in nearby clusters, strongly supporting an evolutionary scenario where the cluster environment acts on newly accreted late-type galaxies, driving a morphological transition towards earlier galaxy types, on timescales of a few Gyr \citep[e.g.][]{BoselliGavazzi06,Donofrio+15}.
However, the question of what is the mechanism responsible for such transition has no unique answer.
The data indicate that the spiral and lenticular fractions have evolved more strongly in lower $\sigma$ (less massive) clusters, while the proportion of ellipticals has remained unchanged \citep{Poggianti+09}. 
This can be explained either by assuming different assembly epochs for clusters of different masses, so that the peak of the morphological transformation would occur earlier (i.e., at epochs not probed by the data) in more massive halos, or by invoking dynamical and/or environmental mechanisms that act preferentially in low-mass clusters or groups.

Amongst the various Hubble types, lenticulars have been historically identified as a transition type between spirals and ellipticals \citep[e.g.][]{Sandage+61}, and have been the focus of a number of observational and theoretical studies aimed at establishing their possible formation pathways. 
The main channels proposed involve the removal and/or the consumption of the gas reservoir (interstellar and circumgalactic medium) in a star-forming spiral, with subsequent quenching of its star formation activity. 
This can be achieved via different processes, with some being more frequent than others depending on the environment.
In small groups or in the field, mergers can dramatically affect the spiral's gas reservoir, causing it to be either flung outward or consumed by in a central burst of star formation, leading to the creation of a classical bulge \citep{Bekki98, Arnold+11, Querejeta+15}.
Gravitational interactions with companions can also promote the formation of a bar in the disc, producing in turn a pseudo-bulge \citep[e.g.][]{Athanassoula05}.
The resulting gas-less system dissipates the original spiral structure through disc instabilities \citep[e.g.][]{Eliche-Moral+13,Saha+18}, producing a smooth, quenched disc surrounding a central bulge.
In the dense environment of a cluster, where galaxy mergers are hampered by the increased relative speed and therefore shorter interaction times \citep[e.g.][]{MakinoHut97}, quenching can be promoted by the gaseous stripping due to hydrodynamical or gravitational forces \citep[e.g.][]{Vulcani+10,Boselli+16, Vulcani+20}.
Ram pressure stripping exerted by the intracluster medium \citep{GunnGott72,Boselli+22} is widely recognised as the most efficient mechanism for gas removal \citep[e.g.][]{Tonnesen+07,Jaffe+15}, and galaxies actively undergoing this process are routinely observed in many clusters, with the most striking instances of these being commonly referred to as `jellyfish' \citep{Smith+10b, Ebeling+14, Poggianti+16, Poggianti+17}.
Gravitational effects such as tidal stripping by the host halo and harassment caused by repeated galaxy-galaxy interactions may also play a role \citep{Moore+96, BoselliGavazzi06, Marasco+16}.

Crucially, different formation mechanisms are expected to leave distinct signature on the lenticular kinematics: systems formed by processes affecting the gas reservoir alone \citep[ram pressure stripping and starvation in clusters, e.g.][]{Larson+80} will maintain the high rotational support and specific angular momentum $j_\star$ typical of spirals, whereas those produced by mechanisms affecting the stars (mergers in the field and groups) will feature lower $v/\sigma$ and $j_\star$ at given stellar mass $M_\star$.
This scenario is fully supported by cosmological hydrodynamical simulations in the $\Lambda$CDM framework \citep{Deeley+21}, and can be tested observationally using kinematic data from integral field spectroscopy (IFS) facilities.
However, the observational picture built so far is rather blurry, with different studies providing somewhat contrasting results.
\citet{Rizzo+18} studied the stellar kinematics of $10$ field/group lenticulars using the Calar Alto Legacy Integral Field Area (CALIFA) survey \citep{Sanchez+12}, finding for 8/10 systems the same $j_\star$-$M_\star$ relation of spiral galaxies \citep[e.g.][]{Posti+18,ManceraPina+21}, indicating that the merger channels is subdominant even in low-density environments.
Using IFS data for $\sim250$ lenticulars from the MaNGA survey \citep{Bundy+15}, \citet{Fraser-McKelvie+18} found that the stellar population properties which are more likely related to the formation channels vary primarily with the galaxy $M_\star$, and do not depend on the environment at all.
Using a larger sample of 329 lenticulars from the SAMI \citep{Bryant+15} and MaNGA surveys, \citet{Coccato+22} found instead a strong environment dependence, supporting the theoretical expectation that the merging channel is dominant in the field and the stripping channel becomes progressively more important with increasing environment density.
Clearly, while stellar kinematics add crucial information to the analysis, more work is needed to clarify the picture.

In this work, we approach the study of the morphological transformation driven by the quenching of star formation from a different perspective.
The assumption underlying our investigation is that the evolution of a quenched galaxy is driven by two separate processes.
The first is the collection of gravitational mechanisms that affect the galaxy dynamics, which include both internal processes (sometimes referred to as `secular') and external ones like tidal interactions with external galaxies and, in the case of cluster galaxies, with the dark matter halo of the cluster itself.
We do not treat these mechanisms in this study, but rather focus on the second process, which is the ageing of the stellar populations.
In the absence of ongoing star formation, the bright and blue continuum light produced by young stars fades out on timescales of a few hundreds Myr. 
Since star formation in disc galaxies is mostly concentrated within spiral arms, ageing implies the fading of the spiral structure itself, which leads to a fainter, redder disc and thus to a larger bulge-to-disc ratio. 
The outcome of this process is that of a morphological transformation towards earlier galaxy types, caused not by an actual change in galaxy structure, but rather by a spatial and spectral redistribution of the stellar light.

The main purpose of this work is to build a machinery that quantifies the magnitude and timescales of such morphological change in galaxies subject to different quenching scenarios. 
Here we will be focusing primarily on gas stripping in clusters, but our methodology can be easily applied to other quenching mechanisms.
Although we present simple applications of our method, this work must be thought as preparatory to a more complex, follow-up evolutionary study that aims at answering the following question: to what degree is the ageing of stellar light responsible for the morphological fractions observed in present-day galaxy clusters?

This paper is structured as follows.
In Section \ref{s:method} we describe the dataset and the methods used to simulate the future spectrophotometric evolution of quenched galaxies and assess the variation of their morphological types as a function of time.
An application of our method for different quenching scenarios is presented in Section \ref{s:results}.
In Section \ref{s:discussion} we discuss the limitations of our evolutionary approach and classification scheme, and describe future outlook.
A summary of this study is given in Section \ref{s:conclusions}.

Throughout this paper we adopt a flat $\Lambda$CDM cosmology with $\Omega_{\rm m,0}\!=\!0.3$ and $H_{\rm 0}\!=\!70\kmsMpc$, and a \citet{Chabrier03} initial mass function (IMF).


\section{Method}\label{s:method}
\begin{figure*}
\begin{center}
\includegraphics[width=0.9\textwidth]{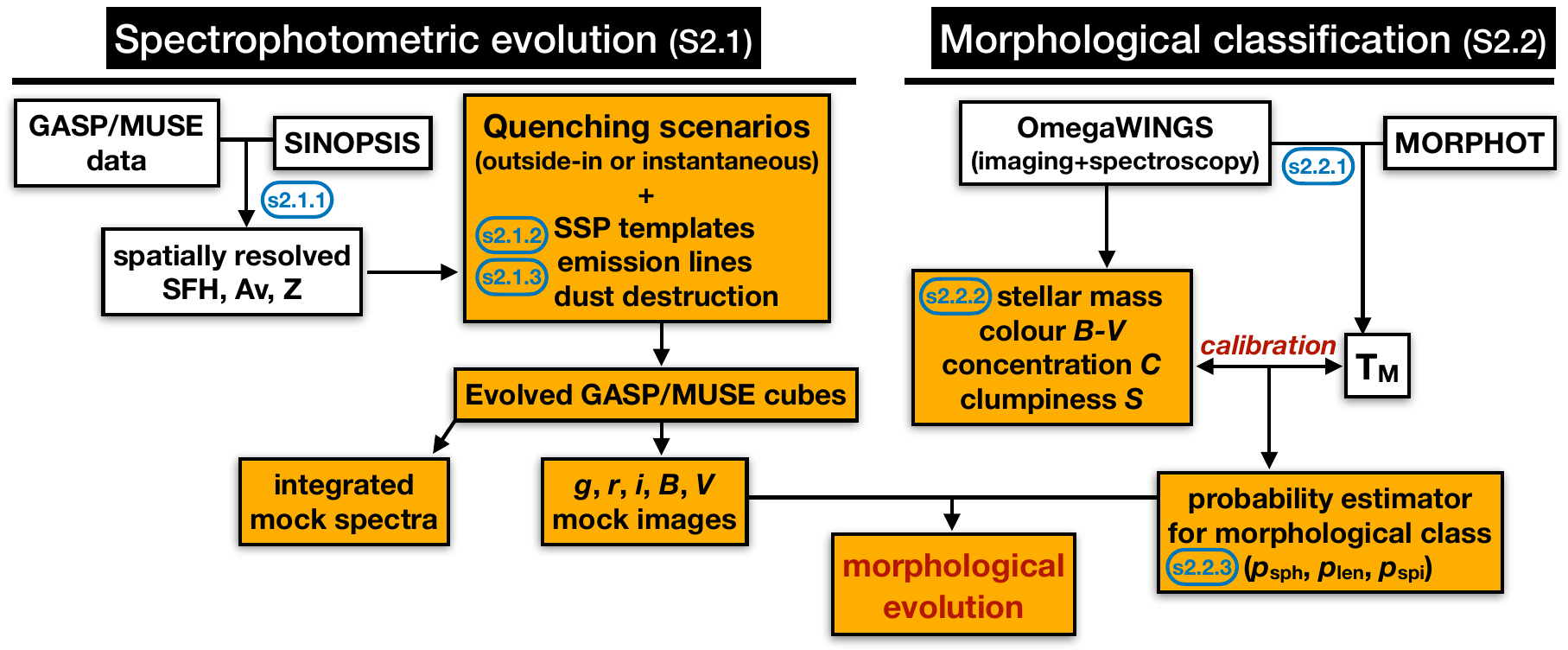}
\caption{Flow chart illustrating our method. White blocks show pre-existing data or analysis products that are used as an input for our method. Yellow blocks show the new analysis products built for this study. The chart indicates also the Sections of the main text that provide details on a given part of the method.
}
\label{f:flow_chart}
\end{center}
\end{figure*}

Our method is divided into two distinct parts.
In the first part (Section \ref{ss:GASPevo}) we show how to derive synthetic spectra and broad-band images for the evolution of any galaxy given a spatially resolved star formation history (SFH) and its future continuation.
In the second part (Section \ref{ss:morphology}) we build a morphology classification scheme, which we calibrate on observed cluster galaxy images, and apply it to the derived synthetic galaxies in order to follow their morphological evolution with time.
We describe in detail our method below, and provide a flow chart (Fig.\,\ref{f:flow_chart}) to guide the reader through its many steps.

\subsection{Spectrophotometric evolution of the GASP sample} \label{ss:GASPevo}
\subsubsection{The GASP sample and its SF history} \label{sss:GASP}
The GAs Stripping Phenomena in galaxies with MUSE (GASP\footnote{\url{http://web.oapd.inaf.it/gasp/index.html}}) project \citep{Poggianti+17} is an integral-field spectroscopic survey with MUSE@VLT aimed at studying gas removal processes in galaxies.
It comprises 114 galaxies at $0.04\!<\!z\!<\!0.07$ with stellar masses $M_\star$ in the range $10^{9.2}-10^{11.5}\msun$, extracted from three surveys that, together, cover the whole range of environmental conditions at low redshift: the WIde-field Nearby Galaxy-clusters survey \citep[WINGS][]{Fasano+06}, the Padova-Millennium Galaxy and Group Catalogue survey \citep[PM2GC][]{Calvi+11}, and the OmegaWINGS survey \citep{Gullieuszik+15}.
GASP is composed by a combination of 94 stripping candidates (64 in clusters and 30 in groups,
filaments, or isolated), which are systems selected to show optical signatures of tails or unilateral debris, and a control sample of 20 cluster and field galaxies without manifest morphological irregularities.

Overall, GASP is an excellent sample for the investigation of the morphology transformation due to stellar population ageing, as it is mostly made by systems that are currently experiencing gas-stripping processes, which will likely lead to SF quenching on short timescales.
However, the variety of environmental processes affecting galaxies in GASP is vast \citep[e.g.][]{Vulcani+21}, especially in low density environments where ram-pressure stripping is less likely to occur.
In this study we focus mainly on what we call the GASP `clean' sample, obtained by removing from the full sample those galaxies that show signs of merging or tidal interactions in their optical images or stellar kinematics \citep[][Poggianti et al. in prep.]{Vulcani+21}, plus systems that do not show emission lines in their MUSE spectra, meaning that they have been passive for at least $\sim20\Myr$ \citep{Vulcani+20}.
This leaves us with 91 galaxies, either unperturbed or at different stages of their stripping process, sparsely distributed in different environments.
We also make use of the GASP control field (CF) sample defined in \citet{Vulcani+18a}, which is a subset of the clean sample made by 15 unperturbed, star-forming galaxies in the field.
Both samples are useful for our investigation.
The former has the advantage of providing better number statistics, although it contains galaxies for which the morphological transformation could be already begun, and progressed to a level that differs from one galaxy to another. 
The CF sample has the advantage of not being affected by this issue, but is much smaller.

Another property that makes GASP the optimal sample for this analysis is the amount of extra information available.
The MUSE data of all GASP galaxies have been modelled with the spectrophotometric fitting code \textsc{sinopsis} \citep{Fritz+07,Fritz+11,Fritz+17} in order to derive spatially resolved stellar metallicity maps, extinction maps for the young ($<20\Myr$) and old ($>20\Myr$) stellar populations separately and, crucially, non-parametric star formation histories in $12$ independent age bins, together with other galaxy properties.
The age bins used by \textsc{sinopsis} are: $0\!-\!2\Myr$, $2\!-\!4\Myr$, $4\!-\!7\Myr$, $7\!-\!20\Myr$, $20\!-\!50\Myr$, $50\!-\!200\Myr$, $0.2\!-\!0.5\Gyr$, $0.5\!-\!1\Gyr$, $1\!-\!3\Gyr$, $3\!-\!5\Gyr$, $5\!-\!10\Gyr$, $10\!-\!13.5\Gyr$.
We refer the reader to \citet{Fritz+17} for further details on the application of \textsc{sinopsis} to the MUSE data of the GASP sample.

The stellar metallicity, extinction and SF history maps determined by \textsc{sinopsis} are used as an input for our spectrophotometric evolution tool.
We stress that, even though we have implemented our routines around the outputs of \textsc{sinopsis}, our procedure is generically applicable to any system for which spatially resolved information for SFH, dust content and stellar metallicity are available.

\subsubsection{From spatially resolved SFH to galaxy spectra}\label{sss:from_SFH_to_spectra} 
We first discuss how synthetic galaxy spectra can be retrieved from a given SFH.
Let us assume a non-parametric SFH that uses $N$ age bins, spaced between each other by a $\Delta t$ dependent on the age, and a constant SFR within each bin.
If the metallicity $Z$ and extinction $A_{\rm V}$ of the stellar population associated with each bin are known, the resulting spectrum $F(\lambda)$ will be given by
\begin{equation}\label{eq:sp_mod}
F(\lambda) = \sum_{i=0}^{N} \left(\dot{M}_i\,\avg{f_i(Z_{\rm i},\lambda)}\,\times 10^{-0.4\,k(\lambda)\,A_{\rm V,i}}\,\Delta t_i \right)
\end{equation}
where the index $i$ indicates a specific age bin, $\dot{M_i}$ is the star formation rate in that bin, $k(\lambda)\equiv A(\lambda)/A_{\rm V}$ is the dust extinction curve, and $\avg{f_i(Z_{\rm i},\lambda)}$ is the mean spectrum produced by a unitary mass single stellar population (SSP) with given metallicity $Z_i$, averaged over the age interval considered.

For consistency with the \textsc{sinopsis} modelling of the GASP MUSE data, in this study we use the extinction law model by \citet{Cardelli+89} with $R_{\rm V}\!=\!3.1$, and the latest set of SSP models by S.\,Charlot and G.\,Bruzual (in prep.\footnote{A description of these models can be found in \citet{Plat+19} and in Appendix A of \citet{Sanchez+22}.})
Emission lines from different atomic species are added to the SSP templates with ages $<20\Myr$, following the prescriptions discussed in S2.3 of \citet{Fritz+17}.
To reduce the number of fitted parameters, \textsc{sinopsis} uses a single value of $Z$ in all age bins of a given spaxel, taken as the value that best reproduces the data.
Also, while separate $A_{\rm V}$ values are determined for the 12 age bins, in the \textsc{sinopsis} outputs used in this work only two values were provided, calculated as the average $A_{\rm V}$ of young ($<20\Myr$, emission lines generating) and old ($>20\Myr$, continuum generating) SSP.
Even with this simplifications, we stress that visual inspection of the model spectra derived via Eq.\,(\ref{eq:sp_mod}) using the \textsc{sinopsis} outputs confirms the excellent agreement between the model and the data \citep[for some examples see][]{Fritz+07}.

\subsubsection{Modelling the evolution}\label{sss:evolution} 
We now start to build the machinery to simulate the spectrophotometic evolution of the GASP galaxies.
Since for each galaxy we will build spatially resolved models for the future continuation of its SFH (see below), our starting ingredient is a map of the current SFR for the galaxy in exam.
Determining this map comes with its complications, for reasons that we now address.
In principle, the SFR map could be taken as $\dot{M}_{i=0}$, that is, the SFR provided by \textsc{sinopsis} corresponding to the most recent age bin ($<2\Myr$).
However, we argue that this choice does not provide a robust description of the recent SFR, given that the MUSE data alone, on which \textsc{sinopsis} has been run, cannot provide a reliable separation of the ages between the different \ha-emitting stellar populations.
A more robust SFR estimate could be inferred by averaging star formation in the latest $t_{\rm SFR}\equiv20\Myr$, which broadly encompass the lifespan of the stellar population responsible for the \ha\ nebular emission line, and correspond to the most recent four \textsc{sinopsis} age bins (see Section \ref{sss:GASP}).
This choice leads though to another complication: if we were replacing the $\dot{M}_i$ for the first four age bins with the corresponding time-averaged value, the spectrum derived via eq.\,(\ref{eq:sp_mod}) would not remain the same.
This occurs because the mass-to-light ratio of SSP varies drastically from one age bin to another, especially for young stars in the age range considered.
We circumvent this issue by imposing that a new SFH, built using a constant (in time) SFR for ages $<t_{\rm SFR}$ and the \textsc{sinopsis} SFH for ages $>t_{\rm SFR}$, leads to a spectrum that is as similar as possible to the original one.
The SFR value that satisfies such condition is derived spaxel by spaxel by fitting the spectrum produced with this new SFH to that derived with the old one.
This procedure provides a new SFH which, by construction, leads to galaxy spectra that are as close as possible to those derived from the \textsc{sinopsis} SFH, and a more robust map for the current SFR, which we use to extrapolate star formation to future times.

Models for the continuation of the SFH are derived as follows.
We consider two simplified scenarios: a gradual outside-in quenching where star formation drops to zero beyond - and proceeds undisturbed within - a circle of progressively diminishing radius, and an instantaneous quenching of the star formation activity over the whole galactic disc.
The former scenario is what we would expect for cluster galaxies subject to ram pressure stripping, with the latter being a limit-case where the stripping is perfectly efficient.
For the former case, the radius of the annulus that separates the star-forming from the quenched regions is modelled as $R(t)\!=\!R_{90}\exp{(-t/t_{\rm strip})}$, where $t_{\rm strip}$ is a characteristic stripping timescale and the $R_{90}$ is defined as the radius encompassing $90\%$ of the reference SFR.
The annulus is centred at the optical centre and projected using the inclination and position angle derived from the optical images (see also Section \ref{ss:morphology}).
This exponential behaviour of the quenching radius leads to a relatively fast quenching of the outer galaxy regions while keeping star formation in the centre intact, in qualitative agreement with the trends observed and expected for ISM stripping by ram pressure.
In this study we test two different values of $t_{\rm strip}$: $0.5$ and $1 \Gyr$. 
Together with the instantaneous quenching case, the scenarios considered broadly encompass the typical timescales of gas removal inferred from  hydrodynamical simulations \citep[e.g.][]{Abadi+99, Tonnesen+09, Marasco+16}.
We note that the stellar mass of the galaxy varies with time depending on the evolution scenario considered and accounting for mass losses due to stellar evolution.

The mock galaxy spectra resulting from such evolution can be derived at any given future time $t$\footnote{We take $t=0$ as the epoch at which the target galaxy is actually observed}.
For any choice of $t$, we build the continuation of the SFH, spaxel by spaxel and according to the chosen scenario, using always additional $20$ age bins to discretise this future time interval. This brings the total number of age bins to $32$, which we found to be an optimal number above which there is no substantial improvement in the synthetic images and spectra derived (see below).
The age bin separation is not constant but follows a power-law function with an index of $3$ (with the first age bin set to $1\Myr$), providing a more refined time resolution for younger ages than for older ones.
As in the original \textsc{sinopsis} run, we keep the metallicity fixed for each spaxel, and use the same $A_{\rm V}$ values determined for the old and young stars.
The novelty that we introduce here is a simple treatment for shock-induced dust destruction, in the absence of future star formation: if $t_{\rm q}$ is the time at which the SFR drops to zero in a given spaxel, we lower the extinction $A_{\rm V}$ at subsequent times by a factor $\exp\left[-(t-t_{\rm q})/t_{\rm dust}\right]$, assuming $t_{\rm dust}\!=\!100\Myr$ \citep[e.g.][]{Jones+94,Jones+96}.
Ignoring dust destruction would make our galaxies redder, accelerating their colour and morphology transformation (Section \ref{s:results}).

The synthetic spectra of the evolved model are determined using Eq.\,(\ref{eq:sp_mod}), after having increased all ages in the SFH by $t$.
In simulating the spectra, we adopt the same wavelength coverage ($4750$--$9350$ \AA) and spectral binning ($1.25$ \AA) of the original MUSE data, which is possible given that quality and extent of the SSP that we use.
We also use the same redshift of the galaxy in exam, determined by the software package KUBEVIZ \citep{Fossati+16} on a spaxel-by-spaxel basis and stored in the \textsc{sinopsis} output.
For the purpose of this paper, we do not store the whole resulting datacube, but rather produce and save mock images in the \emph{gri} SDSS filters and Johnson B and V filters.
All mock images are produced in the observed frame, as if at all times the galaxy was seen at the redshift it is in the observations.
Further discussion on the synthetic images is provided in Sections \ref{ss:results_JO201} and \ref{ss:is_classification_robuts}.
Additionally, to simulate a typical fiber-spectroscopic observation, we store a spectrum integrated within a $2"$-diameter aperture.

\begin{figure*}
\begin{center}
\includegraphics[width=1.0\textwidth]{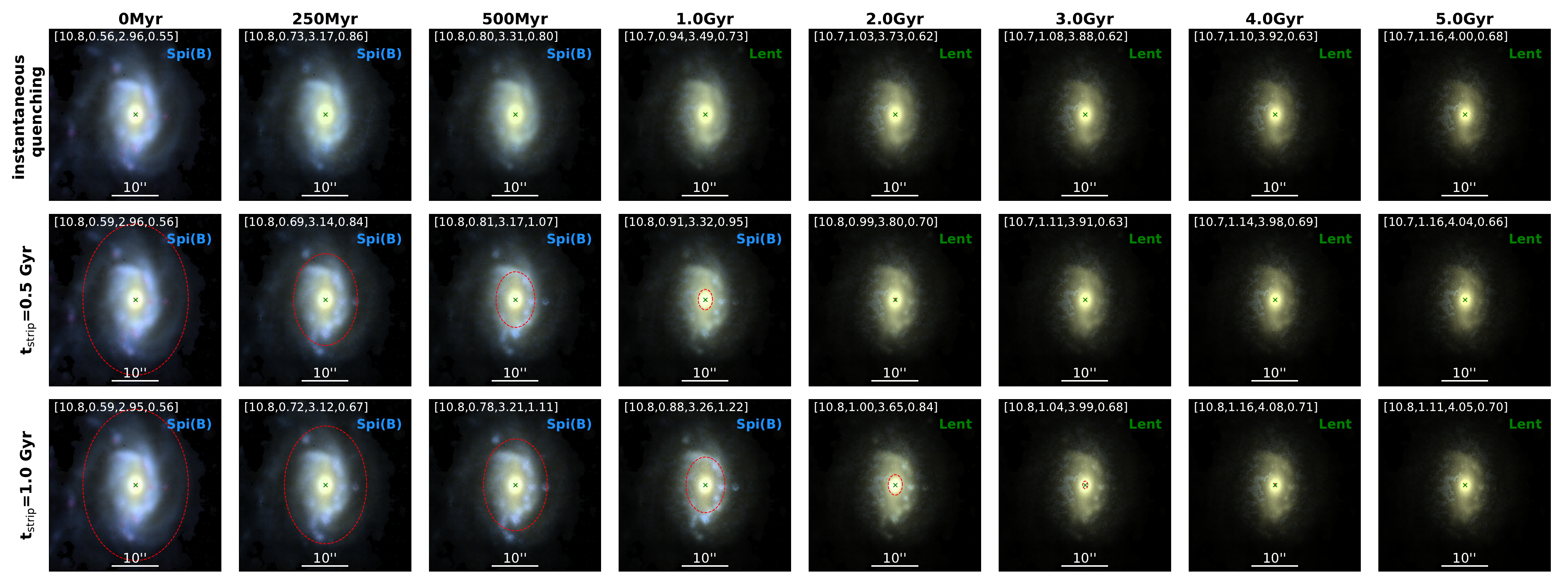}
\caption{Composite \emph{gri} images for the modelled future evolution of galaxy JO201. Each column shows a different future time, ranging from $0$ to $5\Gyr$ as labelled on the top portion of the figure. Each row shows a different evolutionary scenario: instantaneous quenching (top row), or outside-in quenching with stripping timescale of $0.5\Gyr$ and $1\Gyr$ (middle and bottom rows, respectively).
The red-dashed ellipses show the quenching radius for the outside-in quenching scenarios.
The four diagnostic parameters discussed in Section \ref{sss:classif_method}, ($\log_{10}(M_\star/\msun)$, $(B-V)$, $C$, $S$) are listed in the top-left corner of each panel, with the most probable morphological class annotated in the top-right corner. 
}
\label{f:JO201_evo_maps}
\end{center}
\end{figure*}

To illustrate the outcome of our spectrophotometric evolution modelling, we focus on one of the most well-studied `jellyfish' galaxies in the GASP sample, JO201.
JO201 is a late-type galaxy \citep[Sab type, from][]{Fasano+12} moving through the dense intracluster medium of Abell 85 ($z\!=\!0.0557$) at supersonic speeds along the line-of-sight, showing a stellar disc accompanied by large projected tails of \ha-emitting gas that extend out to $\sim100\kpc$ from the galaxy center \citep{Bellhouse+17}.
Signatures of gas stripping are also present in the molecular component traced by CO emission-line observations with APEX \citep{Moretti+18} and ALMA \citep{Moretti+20a,Bacchini+23}.
The kinematic analysis of \citet{Bellhouse+17} suggests that JO201 must so far have lost about half of its gas during infall via ram pressure stripping, in agreement with the \hi-deficiency inferred from JVLA data by \citet{Ramatsoku+20}.
Orbital analysis, together with the study of the stellar populations in the disc and in the tail, indicates a stripping time-scale of $\sim1\Gyr$ \citep{Bellhouse+19,George+19}.
Clearly, JO201 is in the process of losing its gaseous disc and quenching its star formation activity, which makes of it a good test case for our analysis. 

Fig.\,\ref{f:JO201_evo_maps} shows a series of composite \emph{gri} (SDSS-like) images at different times for the future evolution of JO201, for each of the three scenarios proposed.
Clearly, as the galaxy evolves with time (proceeding from left to right in Fig.\,\ref{f:JO201_evo_maps}), it is subject to a number of photometric and structural changes: its disc, initially blue and with a marked spiral arm structure, becomes redder, fainter and smoother, with no clear indication of a spiral arm pattern after $3\!-\!4\Gyr$.
The dimming of the disc is more pronounced in the galaxy outskirts, where the stellar population is younger, leading to a progressive shrinking of the galaxy effective radius down to $\sim70\!-\!80\%$ of the initial value, with small variations depending on the band and quenching scenario considered. 
The morphological transformation occurs in all evolutionary scenarios analysed (different rows in Fig.\,\ref{f:JO201_evo_maps}), but there are minor differences due to star formation proceeding in the inner regions for the outside-in scenarios.
The differences are particularly evident at $t\!=\!1\!-\!2\Gyr$, where the instantaneous quenching scenario provides a galaxy which is already red and smooth, whereas blue and clumpy regions are still visible in the disc for the outside-in quenching cases, especially for longer $t_{\rm strip}$, as expected.

Although only qualitative, this experiment shows the potential of our investigation method: once star formation is halted, JO201 transforms into a lenticular galaxy on timescales that vary depending on the assumed quenching scenario.
The purpose of the next Section is to quantify such transformation. 

\subsection{Morphological classification} \label{ss:morphology}
In order to quantify the morphological evolution of our modelled systems, we need a tool to infer their morphological class.
We stress that in this study we focus on a `classical' morphological classification based on broad-band imaging, which we can model, and not on galaxy kinematics, whose evolution cannot be inferred with our approach.

\subsubsection{The OmegaWINGS sample}\label{sss:omegawings}
As we are modelling galaxies subject to ram pressure stripping from the intracluster medium, our strategy is to calibrate our classification scheme using an observed sample of cluster galaxies for which morphology has been already robustly determined.
The ideal sample for this analysis is that from OmegaWINGS \citep{Gullieuszik+15}.

OmegaWINGS is survey of 46 nearby ($0.04\!<\!z\!<\!0.07$) galaxy clusters and of their peripheries, imaged in the \emph{B} and \emph{V} bands with OmegaCAM \citep{Kuijken11} at the VLT Survey Telescope \citep[VLS;][]{CapaccioliSchipani11} in Cerro Paranal.
The $1$ deg$^2$ FOV camera covers each cluster beyond its virial radius, permitting to image the surrounding infall region, effectively extending the environment range covered by the parent WINGS Survey \citep{Fasano+06}.
Spectroscopic follow-ups for 36 OmegaWINGS clusters were taken with the AAOmega spectrograph \citep{Smith+04,Sharp+06} at the Australian Astronomical Observatory (AAT) for redshift measurements and assignment of cluster memberships \citep{Moretti+17}.

As previously done for WINGS, the morphological classification of OmegaWINGS galaxies has been performed using the MORPHOT package \citep{Fasano+12} by \citet{Vulcani+23}.
MORPHOT exploits 21 morphological diagnostics, computed directly from the \emph{V}-band images, to provide two independent classifications (one based on a maximum likelihood technique, and the other one on a neural network machine, both calibrated on visual inspection), which in turns are used to infer a morphological type $T_{\rm M}$.
Unfortunately, MORPHOT has been specifically designed for the WINGS/OmegaWINGS observations and is not readily applicable to other datasets.
Hence, for our classification we must rely on correlations between $T_{\rm M}$, which is available for the whole OmegaWINGS sample, and an independent set of diagnostics that can be easily computed for our evolved GASP galaxies and for systems in OmegaWINGS.
In other words, we will be using the $T_{\rm M}$ type inferred from a highly complex and refined classification scheme such as MORPHOT, as calibrator for our simpler, but more easily applicable classification approach.

\subsubsection{Photometric and structural properties of OmegaWINGS galaxies}\label{sss:photo_struct}
We now build an optimal set of diagnostics for our classification scheme.
We first extract $1'\times1'$ cutouts (corresponding to the MUSE field of view) in \emph{B} and \emph{V} bands for all OmegaWINGS galaxies with available $T_{\rm M}$ measurements.
These images are then processed with an updated version of the image analysis software package presented in \citet{Marasco+23} \citep[Appendix A; see also][]{Marasco+19a}.
In its original version, this software computes photometry by extracting the cumulative light profile from sky-subtracted images, after the (automatic) removal of contamination from point-like sources such as foreground stars and background galaxies.
The calibration of the resulting photometry in \emph{B} and \emph{V} bands is done for each cluster separately, using Table A.1 in \citet{Gullieuszik+15}.
After calibration, there is excellent agreement between our photometric measurements and those of \citet{Gullieuszik+15}, with rms scatter of $0.06$ and $0.05$ mag (well within the photometric errors) in the \emph{B} and \emph{V} bands, respectively.

With respect to the routines presented in \citet{Marasco+23}, we have implemented two main improvements.
The first is the automatic estimate of the galaxy geometry, namely of its centre, inclination, position angle, and maximum extent $R_{\rm max}$, which are used by the photometric routine to separate the `sky' from the`galaxy' regions, as discussed in \citet{Marasco+23}.
The galaxy centre is computed iteratively as the light-weighted centroid in apertures with progressively diminishing sizes.
The other geometrical quantities are determined using the 2nd-order moments of the pixel intensity distribution, as in the \textsc{SExtractor} package \citep{sextractor}, but using only the pixels within a mask, which is built via a smooth-and-clip technique.
$R_{\rm max}$, in particular, is set as $5\times A$, where $A$ is the length of the galaxy semi-major axis determined as the maximum spatial dispersion of the object profile along any direction, which can be computed from the image 2nd-order moments \footnote{In \textsc{SExtractor}, the default choice for $R_{\rm max}$ is $3\times A$, but here we prefer to adopt a more conservative separation between the galaxy and the sky region of the image.}.
Further information on the \textsc{SExtractor} routines can be found at the software webpage\footnote{\url{https://sextractor.readthedocs.io/en/latest/Position.html}.}.
The second improvement is the computation of the so-called CAS structural quantities - concentration (C), asymmetry (A) and clumpiness (S) - following \citet{Conselice+00} and \citet{Conselice03}, which provide a non-parametric characterisation of the galaxy structure in a given band.
In short, these parameters quantify the degree by which the galaxy light is concentrated towards the innermost regions (C), is asymmetrically distributed with respect the the system's centre (A), and has an overall clumpy appearance (S).

The outcome of our OmegaWINGS analysis is a catalogue with the geometric, photometric and CAS structural parameters for all spectroscopically confirmed cluster members with assigned $T_{\rm M}$, for the \emph{B} and \emph{V} bands separately.
Stellar masses $M_{\star}$ for cluster members are
computed by \citet{Vulcani+22} following \citet{Bell+01} and exploiting the correlation between stellar mass-to-light ratio and optical colors of the integrated stellar populations.
Further information on stellar mass computations for WINGS/OmegaWINGS cluster members are provided by \citet{Vulcani+11} and in Section 2 of \citet{Vulcani+22}.
From this sample, we remove systems with inclination $\!>\!70\de$, for which both colour and structure can be heavily influenced by dust.
Additionally, a completeness correction weight $w\!\equiv\!1/R$ (with $0\!<\!R\!\leq\!1$) is associated with each OmegaWINGS cluster member, where $R$ is derived as the ratio of number of spectra yielding a redshift to the total number of galaxies in the parent photometric catalog, calculated both as a function of V magnitude and radial projected distance from the brightest cluster galaxy \citep{Paccagnella+16, Moretti+17}.
We use these completeness weights for our classification, as we discuss in Section \ref{sss:classif_method}.
In our analysis we exclude galaxies with $R\!<\!0.2$ as they are excessively under-represented in the observed sample, so their large $w$ can lead to spurious results.
This leaves us with $7719$ (unweighted) galaxies with V-band magnitude $\!<\!20$ and $M_\star\gtrsim10^{8.5}\msun$ which we use to build our morphology classification method.

\subsubsection{A 4D morphology classification method}\label{sss:classif_method}
We begin by identifying the parameters in our catalogue that are more strongly dependent on the MORPHOT type $T_{\rm M}$.
We identify four main quantities, ranked by the modulus of their Pearson (P) correlation coefficient: $C$ ($P\!=\!-0.71$), $S$ ($P\!=\!0.60$), the $(B-V)$ (observed-frame) colour ($P\!=\!-0.50$), and $M_\star$ ($P\!=\!-0.30$).
This is expected, given that the different morphological types are well separated in the color-mass plane \citep[e.g.][see also Fig.\,\ref{f:color_mass_diagram} below]{Tully+82,Baldry+04,Schawinski+14,Powell+17}, and that earlier galaxy types have typically higher C (or, equivalently, higher Sérsic index) and an overall smoother appearance (that is, lower S) due to the lack of spiral arms with ongoing star formation \citep[see also][]{Driver+06}.
Correlation with other quantities, such as the asymmetry $A$, are poorer, thus we prefer to limit our calibration method to this four-dimensional ($M_{\star}$, $C$, $S$, $(B-V)$) parameter space, considering also that the number of galaxies available would not permit a more refined classification method.
Fig.\,\ref{f:corner} in Appendix \ref{app:pspi_pell} provides corner plots showing the relations between different parameter couples for galaxies in OmegaWINGS.

The use of $(B-V)$ colour in the observed-frame rather then in the rest-frame is justified by the fact that the redshift range spanned by OmegaWINGS cluster galaxies is relatively narrow.
Visual inspection of the $M_\star$-$(B-V)$ plane shows no substantial difference between galaxies in different quartiles of the redshift distribution.
Also, using mock galaxy catalogues from the `Spectro-Photometric Realisations of IR-Selected Targets at all-z' (SPRITZ) simulations \citep{Bisigello+21}, we find that variations in the $k$ correction to the ($B-V$) within the redshift range $0.04\!<\!z\!<\!0.07$ are always $\!<\!0.1$ dex for all galaxy types.

As a next step, we further simplify our classification scheme by assigning each galaxy in our OmegaWINGS catalogue to one of these three `broad' morphological classes: ellipticals (Ell) for $-5.5\!<\!T_{\rm M}<-4.25$ ($21\%$), lenticulars (Len) for $-4.25\!\leq\!T_{\rm M}\!\leq\!0$ ($43\%$), and spirals (Spi) for $T_{\rm M} \!>\!0$ ($36\%$).
The $T_{\rm M}$ ranges defining these classes have been originally derived by \citet{Fasano+12}, and are based on the comparison between the $T_{\rm M}$ values of $979$ galaxies, randomly selected from the WINGS survey, and their morphological type determined by visual inspection of their V-band images\footnote{The ranges adopted here follow those used by \citet{Vulcani+23}, which are slightly different than the original ones from \citet{Fasano+12}.}. 
The OmegaWINGS catalogue is used to derive the probability that any given galaxy with measured $\Vec{x}\!=\!(M'_{\star},C',S',(B-V)')$ belongs to each of these morphological classes.
This is computed as the (completeness-weighted) fraction of Ell, Len and Spi in the first $\mathcal{N}$ galaxies `closer' (see below) to position $\Vec{x}$ in the OmegaWINGS catalogue:
\begin{equation}\label{eq:probability}
    \Vec{p}(\Vec{x})\equiv(p_{\rm ell},p_{\rm len},p_{\rm spi})=\frac{1}{\sum_{i=0}^{\mathcal{N}}(w_i)}\,\left(\sum_{i=0}^{\mathcal{N}_{\rm ell}} w_i, \sum_{i=0}^{\mathcal{N}_{\rm len}} w_i, \sum_{i=0}^{\mathcal{N}_{\rm spi}} w_i\right)
\end{equation}
where $w_i$ are the completeness weights (Section \ref{sss:photo_struct}), $(p_{\rm ell}\!+\!p_{\rm len}\!+\!p_{\rm spi})\!=\!1$ and the three sums in the right-most side of Eq.\,(\ref{eq:probability}) are limited to the three morphology classes considered.
The proximity of OmegaWINGS galaxies to $\Vec{x}$ in the 4D space considered is determined using an Euclidean distance $d$ defined as
\begin{multline}\label{eq:metric}
    d(\Vec{x}) = \biggl[\left(\frac{\log_{10}M_\star-\log_{10}M'_\star}{\delta\log_{10}M_\star}\right)^2 + \left(\frac{C-C'}{\delta C}\right)^2 + \left(\frac{S-S'}{\delta S}\right)^2 +\\ \left(\frac{(B-V)-(B-V)'}{\delta(B-V)}\right)^2 \biggr]^{1/2}\,.
\end{multline}
The various $\delta$ terms at the denominators of equation (\ref{eq:metric}) ensure that the 4D parameter space is normalised so that the dynamic range spanned by each `direction' is similar, and are computed as the difference between the $95$th and $5$th percentiles for each quantity in the OmegaWINGS sample.
The choice of $\mathcal{N}$ (the number of OmegaWINGS systems used for the `local' fraction computations) is somewhat arbitrary, but we find that values between $20$ and $100$ ensure a smooth characterisation of $\Vec{p}(\Vec{x})$ and lead to similar results, thus we adopt $\mathcal{N}\!=\!50$.

As OmegaWINGS galaxies do not populate the 4D space uniformly, it may happen that the galaxy number density around a given $\Vec{x}$ is too small to be informative of the morphological fractions at that location.
If $d(\Vec{x})_{|\mathcal{N}}$ is the distance of the furthest $\mathcal{N}$-th galaxy from  position $\Vec{x}$ computed via eq.\,(\ref{eq:metric}), we set a lower limit to the galaxy number density $n\equiv \mathcal{N}/\frac{4}{3}\,\pi\, d^3(\Vec{x})_{|M}$ above which the calculation of $\Vec{p}(\Vec{x})$ is permitted.
Regions of the parameter space where $n$ is below this threshold, which we set to $30$ by visual inspection of the probability distributions (see below), will output an `unclassified' galaxy type.
We stress that, even though our choices for $\mathcal{N}$ and the threshold in $n$ are somewhat arbitrary, experiments with different values did not result in substantial differences with respect to the findings presented in Section \ref{s:results}.

\begin{figure*}
\begin{center}
\includegraphics[width=\textwidth]{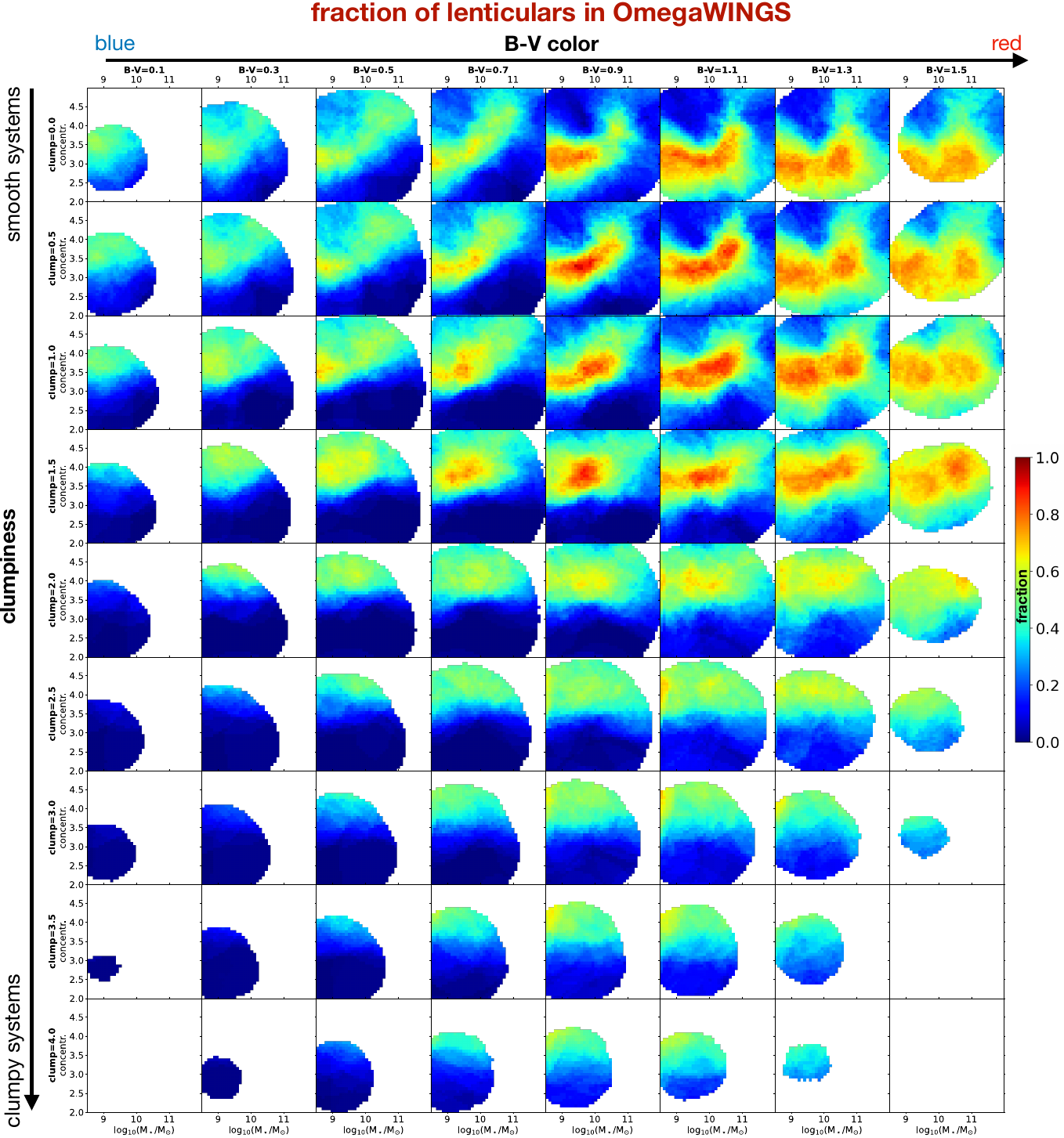}
\caption{Fraction of lenticular galaxies (defined as systems with $-4.25\!<\!T_{\rm M}\!\leq\!0$) in the OmegaWINGS sample as a function of stellar mass $M_\star$, $(B-V)$ colour, concentration $C$ and clumpiness $S$. Colour-coding in individual panels show the class fraction in the $C$-$M_\star$ plane. $(B-V)$ increases from left to right, $C$ increases from top to bottom. Only spectroscopically confirmed cluster members with inclination $<70\de$ and completeness parameters $k\!\geq\!0.2$ have been used. Empty areas show the `unclassified' region of the parameter space (see text).}
\label{f:OW_len}
\end{center}
\end{figure*}

To illustrate the probabilities computed with our method, in Fig.\,\ref{f:OW_len} we show the $p_{\rm len}(\Vec{x})$, that is the fraction of lenticular galaxies in the OmegaWINGS sample computed via Eq.\,(\ref{eq:probability}).
Each panel shows the $p_{\rm len}$ in the $C$-$\log_{10}{M_\star}$ plane at given color $(B-V)$ and clumpiness $S$, with the $B-V$ increasing from left to right and $S$ increasing from top to bottom. 
The empty regions of each panel show the location of unclassified ($n<30$) objects, following the definition of above.
Clearly, the probability that a galaxy is classified by MORPHOT as a lenticular is maximised for red, smooth systems with intermediate $M_\star$ and $C$.
Vice-versa, clumpy, blue galaxies are never lenticulars, unless they have $C\!>\!3.5$. 
Low-$C$ galaxies can be lenticulars only if they are very smooth ($S<1.5$) and red.
Overall, the trends shown in Fig.\,\ref{f:OW_len} provide a strong indication that all $4$ parameters considered play a role in setting the morphological class of a galaxy.
Figures analogous to Fig.\,\ref{f:OW_len} for ellipticals and spirals are shown in Appendix \ref{app:pspi_pell}.

\subsubsection{Classification strategy and error estimates}\label{sss:classif_uncertainties}
\begin{figure}
\begin{center}
\includegraphics[width=0.49\textwidth]{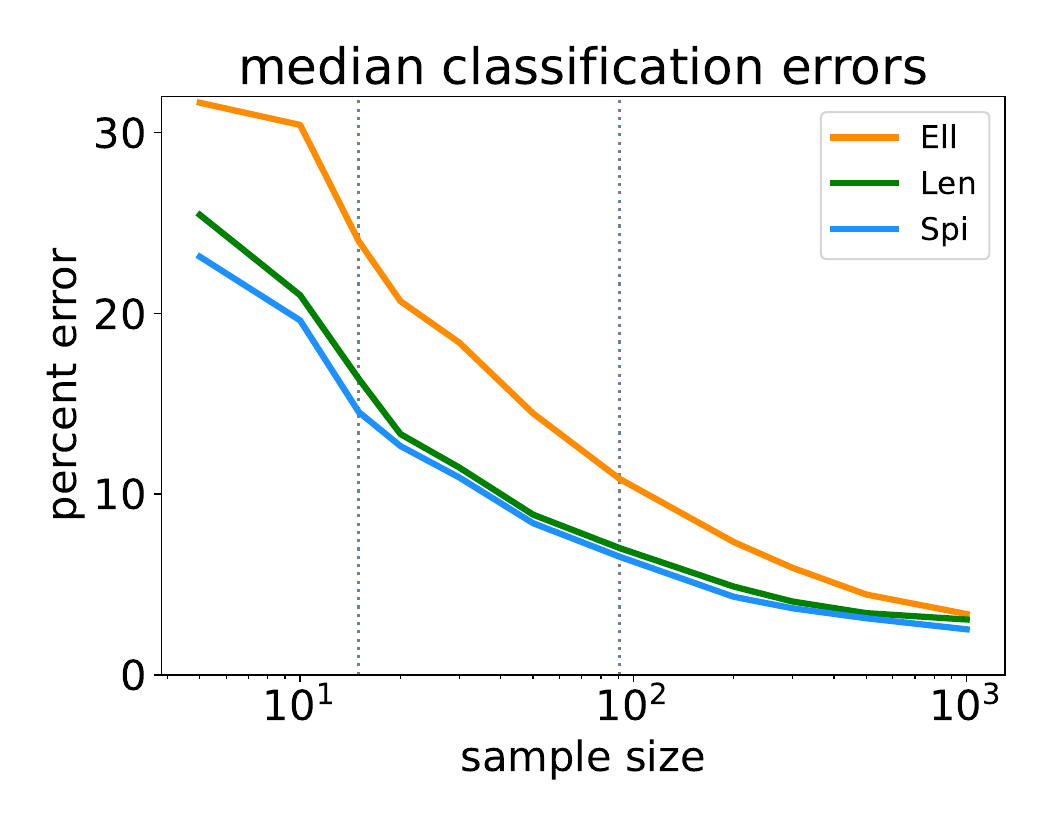}
\caption{Median uncertainties associated with our statistical classification method as a function of the sample size. Uncertainties are shown separately for each morphological class. The dotted vertical lines mark the size of the GASP clean sample (91) and of its control sample of field galaxies (15).}
\label{f:classification_error}
\end{center}
\end{figure}

We now discuss how the probabilities defined by Eq.\,(\ref{eq:probability}) can be used for the morphological classification of the `evolved' GASP galaxy sample built in Section \ref{sss:evolution}.
The simplest approach is that of an individual classification, performed galaxy-by-galaxy by computing Eq.\,(\ref{eq:probability}) and assigning each system to its most probable class.
The classes reported in the top-right corners of the various panels in Fig.\,\ref{f:JO201_evo_maps} have been assigned with this method.
However, while this method is useful to characterise the morphology of a single system, its downside is that regions of the parameters space for which $p_{\rm ell}$, $p_{\rm len}$ and $p_{\rm spi}$ are similar will probably lead to a misclassification.
A more robust approach is that of a statistical classification, where the probabilities computed via Eq.\,(\ref{eq:probability}) on the the entire sample are added together to infer the class distribution of the galaxy population as a whole.
This is the classification method that we adopt in this study.

The accuracy of our statistical classification method is expected to vary with the size $N$ of the galaxy sample in exam, improving for larger $N$.
Low number statistics will affect the classification of small samples: an extreme case is a sample with $N\!=\!1$, for which we find the same problem discussed above for the `individual' classification method.
On the other hand, if we were trying to re-classify the entire OmegaWINGS sample with our method, we would expect to recover the fractions of Ell, Len and Spi given by MORPHOT with almost perfect accuracy, given that the method is calibrated on the whole OmegaWINGS catalogue to begin with.
These considerations suggest that we can quantify the accuracy of our classification scheme as a function of the sample size by considering $N$ randomly extracted galaxies from the OmegaWINGS sample itself, and comparing the `true' and the predicted fractions of each morphological class.
This test also allows us to assess how much information is lost by passing from the 21-dimensional parameter space explored by MORPHOT to the 4D parameter space used in this study.

In practice, we proceed as follows. 
For a given $N$, we create $1000$ different subsamples of $N$ galaxies randomly extracted from OmegaWINGS.
For each subsample we compute the `true' fractions of each class ($f_{\rm ell}$, $f_{\rm len}$, $f_{\rm spi}$, based on OmegaWINGS MORPHOT classifications) and those inferred from our statistical method ($f'_{\rm ell}$, $f'_{\rm len}$, $f'_{\rm spi}$).
We define the uncertainty of our classification scheme for a sample of size $N$ as the median of the distribution of $|f_{\rm i}-f'_{\rm i}|/f_{\rm i}$, with the index $i$ indicating the class in exam.

The results of this analysis are presented in Fig.\,\ref{f:classification_error}, which shows the median uncertainty as a function of $N$ for each class separately.
The trend with $N$ is evident: typical uncertainties in the morphology fraction range from $25-30\%$ for samples of a few galaxies, to a few $\%$ for samples of $10^3$ objects.
The elliptical fractions have systematically larger uncertainties, possibly because this class is the least represented in OmegaWINGS ($21\%$ Ell, $43\%$ Len, $36\%$ Spi), whereas we find similar errors for the other two classes.
The vertical dotted lines in Fig.\,\ref{f:classification_error} mark the sizes of the GASP clean sample (91 objects) and of the CF sample (15 objects) that will be analysed in Sections \ref{ss:results_GASP} and \ref{ss:results_subGASP}, and for which we use the uncertainties computed here.

\section{Results}\label{s:results}
\subsection{Application to JO201}\label{ss:results_JO201}
\begin{figure}
\begin{center}
\includegraphics[width=0.49\textwidth]{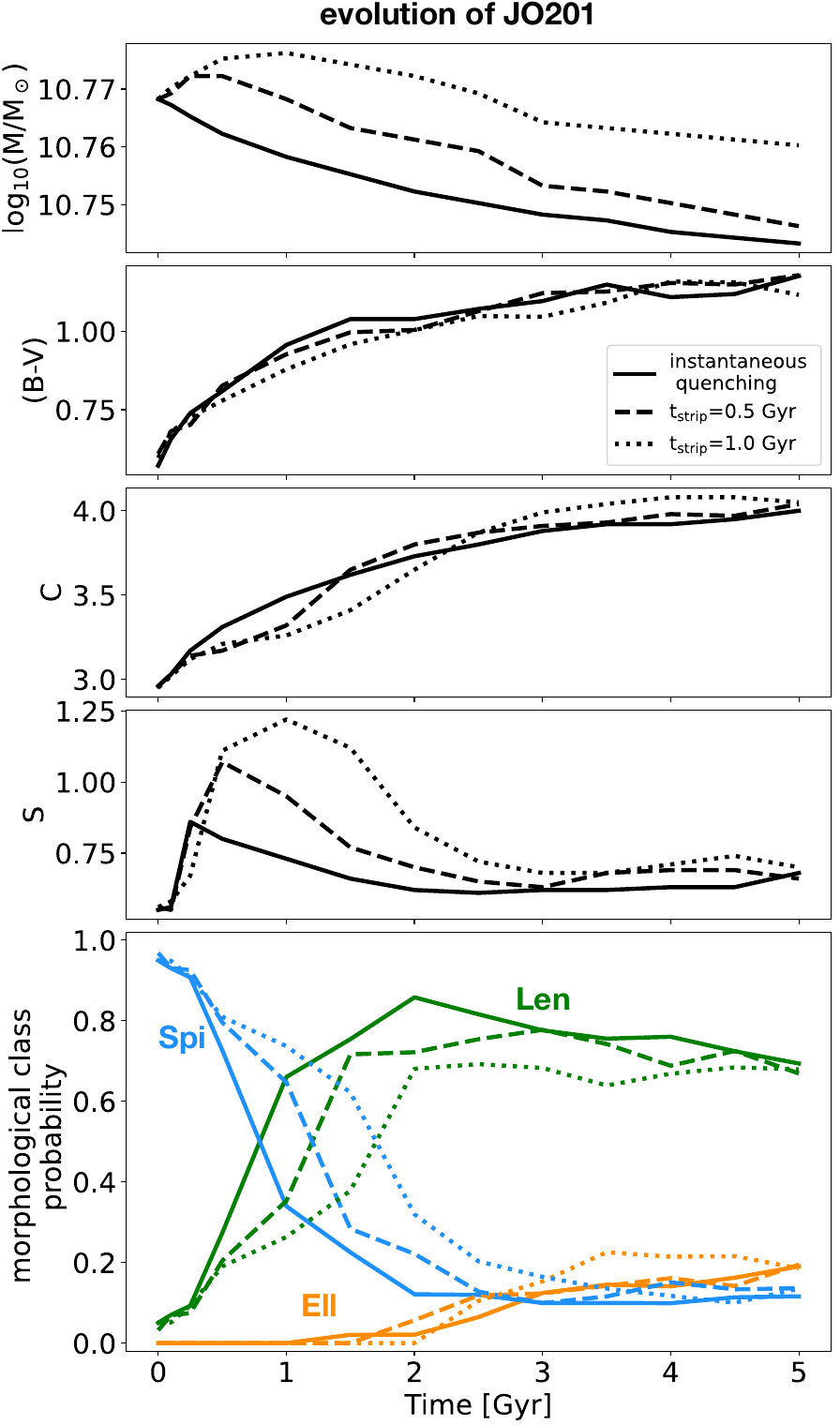}
\caption{Time-evolution for the morphological diagnostic parameters in JO201. From top to bottom: stellar mass $\log_{10}(M_\star/\msun)$, colour $(B-V)$, concentration $C$, clumpiness $S$. 
The lowermost panel shows the time-evolution of the morphology class probabilities computed via Eq.\,(\ref{eq:probability}), using separate colors for each class (blue for Spi, green for Len, orange for Ell).
In all panels, different lines styles are used for the three evolutionary scenarios considered: instantaneous quenching (solid), outside-in quenching with $t_{\rm strip}\!=\!0.5\Gyr$ (dashed) and with $t_{\rm strip}\!=\!1\Gyr$ (dotted).}
\label{f:JO201_evo_diagnostic}
\end{center}
\end{figure}

As a test case, we now go back to the evolution of JO201 shown in Fig.\,\ref{f:JO201_evo_maps}
and compute Eq.\,(\ref{eq:probability}) for all timesteps and quenching scenarios.
The resulting probabilities are shown in Fig.\,\ref{f:JO201_evo_diagnostic}, together with the time evolution of the four morphology diagnostics.

We can now quantify the structural and photometric evolution that we have discussed only qualitatively in Section \ref{sss:evolution}.
Clearly, both the colour $(B-V)$ and the concentration $C$ grow steadily with time, which is a clear sign of a transformation towards an earlier galaxy type.
As expected, these growths are initially faster for the instantaneous quenching scenario, and slower for the outside-in quenching with the largest $t_{\rm strip}$.
Larger $t_{\rm strip}$ values also lead to a marginally higher $C$ at later times, due to the increased central mass growth.
However, we stress that for all scenarios considered the growth of $C$ is largely dominated by the fading of the outer disc light, rather than by the actual mass growth in the central regions (bulge).

Stellar masses at different timesteps are derived by integrating the SFH.
We include stellar remnants in the $M_\star$ budget, and subtract mass losses due to stellar winds and supernovae\footnote{We note that stellar masses computed with \textsc{SINOPSIS} from the MUSE data are slightly larger than those obtained for the OmegaWINGS sample by \citet{Vulcani+22} from the photometry. To be fully compatible and allow a direct comparison, we have applied a correction to the $M_\star$ of each GASP system, derived by fitting the relation between the $M_\star$ derived with the two methods with a straight line.
We determine the correction at $t=0$ and then apply it at all timesteps.
This correction is small: in JO201 it is just $0.04$ dex, and ranges from $0.01$ to the $0.24$ dex for the rest of the GASP sample.}.
As shown in the top panel of Fig.\,\ref{f:JO201_evo_diagnostic}, $M_\star$ in JO201 decreases monotonically with time for the instantaneous quenching scenario, due to the fact that the mass in stars decreases as stars die out.
Since SF proceeds longer for the outside-in quenching cases, the $M_\star$ features an initial phase of growth that lasts up to $\sim2\times t_{\rm strip}$, followed by the same decay of the instantaneous quenching case.
We notice that, in all scenarios considered, variation in $M_\star$ are only marginal ($\!<\!10\%$) during the $5\Gyr$ period considered, and would be even lower if we had considered a bottom-heavy IMF \citep[e.g.][]{Salpeter55} given that more mass would be locked in long-lived stars.

The trend of the clumpiness $S$ varies substantially from one scenario to another: while $S$ is about constant for the instantaneous quenching, it grows by a factor $2\!-\!3$ in the other scenarios during the initial $\sim2\times t_{\rm strip}$ yr, decreasing back to the original values afterwards. 
This growth is driven by SF activity proceeding only in those regions where recent SF was already present, which contributes to enhance substructures in the disc before quenching occurs.
The $S$ enhancement for the outside-in quenching cases can be seen visually in Fig.\,\ref{f:JO201_evo_diagnostic} at $t\!=\!0.5-2\Gyr$, where the outer disc regions are populated by multiple blue stellar clumps.
We discuss further on how the lack of dynamical evolution in our models impact the estimate of $S$ in Sections \ref{ss:results_subGASP} and \ref{ss:limitations}.
We also stress that, amongst the four parameters considered, $S$ is the most sensitive to the spatial resolution and overall quality of our images.
This caveat is addressed further in Section \ref{ss:is_classification_robuts}.

The bottom panel of Fig.\,\ref{f:JO201_evo_diagnostic} shows the evolution of the morphology class probabilities computed with Eq.\,(\ref{eq:probability}).
All quenching scenarios follow a similar trend: while initially JO201 is undoubtedly a spiral galaxy, as the evolution proceeds, the probability that the system is classified as a spiral drops in favour of the lenticular and, to a less extent, the elliptical class.
The transition from spiral to lenticular as the most probable class occurs on different times depending on the scenario considered, being faster ($0.9\Gyr$) for the instantaneous quenching case and slower ($1.8\Gyr$) for the outside in stripping case with $t_{\rm strip}\!=\!1\Gyr$.
The latter scenario also features a larger probability for a Ell classification, which is caused by the slightly higher values of $C$ reached by this model.

We stress that the application of Eq.\,(\ref{eq:probability}) on a generic galaxy image must be done with caution, as our method is calibrated on a well defined set of galaxy images with given spatial resolution and sensitivity.
In applying our classification to the synthetic GASP images, we must make sure that their resolution and signal-to-noise are compatible with those of OmegaWINGS observations. 
Since both MUSE and OmegaCAM are seeing limited instruments and have similar pixel sizes, we do not adjust the spatial resolution of our synthetic images, which inherit the MUSE resolution from \textsc{sinopsis}. 
However, we inject into them the same level of background noise (assumed Gaussian, for simplicity) of the OmegaWINGS data, computed for the $B$- and $V$- bands separately.
For GASP galaxies in OmegaWINGS, this is determined galaxy-by-galaxy by computing the rms noise in the background-subtracted images in regions outside the galaxy, while for the remaining systems we use the median of the noise distribution determined for the others.
Further consideration on our classification method are given in Section \ref{ss:is_classification_robuts}.

Overall, our analysis indicates that the quenching of star formation in a grand design spiral galaxy like JO201 will induce a morphological transformation towards the lenticular class (and, to a lesser extent, towards the elliptical class, especially for long stripping timescales) on timescales that vary between $1$ and $2\Gyr$ from the time the quenching initiated, depending on the details of the quenching mechanism itself.

\subsection{Morphological evolution in GASP}\label{ss:results_GASP}

\begin{figure*}
\begin{center}
\includegraphics[width=0.49\textwidth]{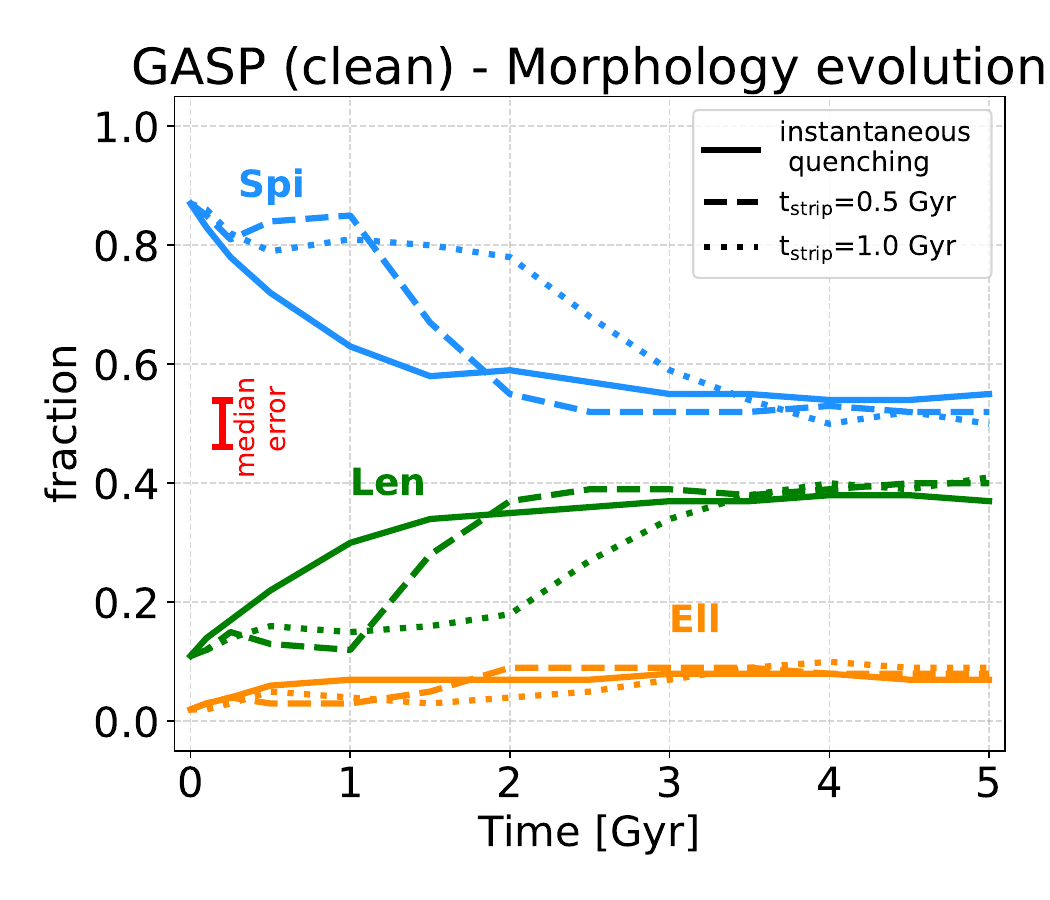}
\includegraphics[width=0.49\textwidth]{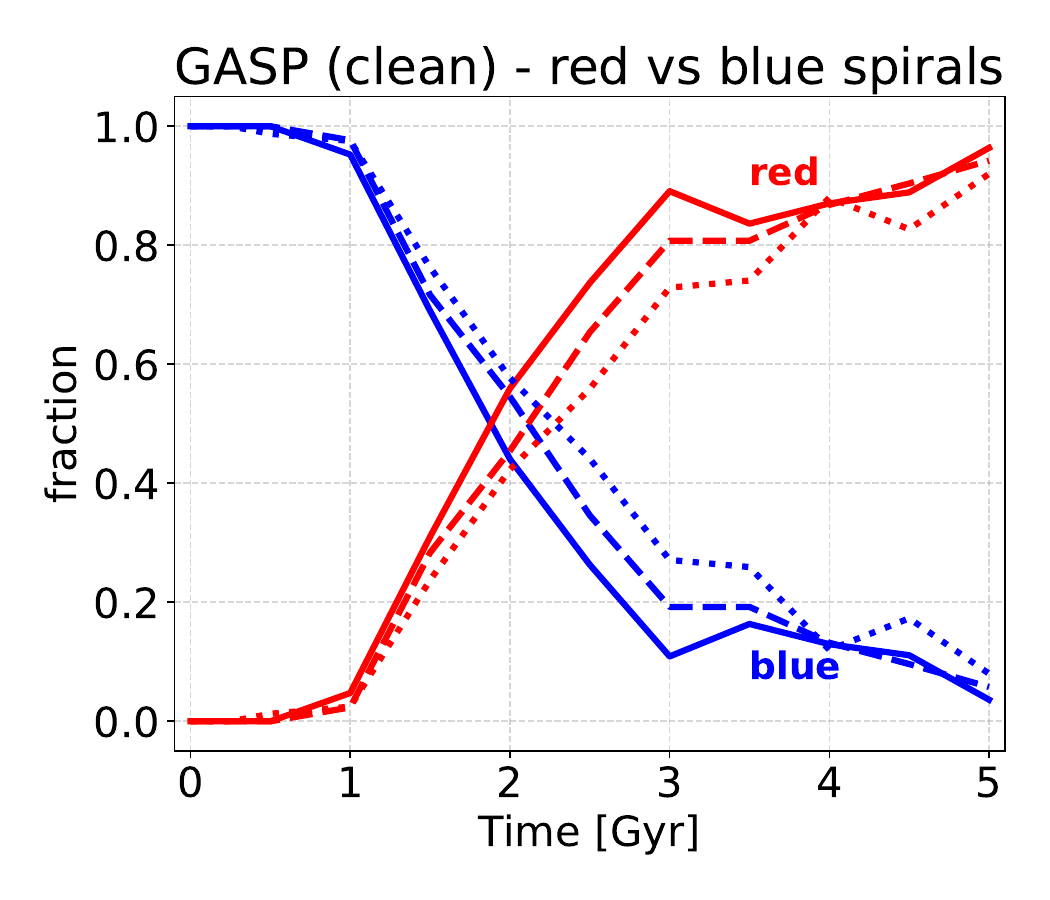}
\caption{Morphological evolution of the GASP `clean' sample of 91 galaxies for the three quenching scenarios considered (solid lines: instantaneous; dashed lines: outside-in with $t_{\rm strip}\!=\!0.5\Gyr$; dotted lines: outside-in with $t_{\rm strip}\!=\!1.0\Gyr$). The \emph{left panel} shows the evolution of the fraction of GASP galaxies of a given morphological class (blue: spirals, green: lenticulars, orange: ellipticals), determined with the statistical method discussed in Section \ref{sss:classif_method}. The red error-bar shows the typical uncertainty in the fraction estimates. The \emph{right panel} shows the evolution of the red and blue spiral fractions, normalised by the total number of spiral galaxies at a given time.}
\label{f:morph_evo_all_stat}
\end{center}
\end{figure*}

\begin{figure}
\begin{center}
\includegraphics[width=0.49\textwidth]{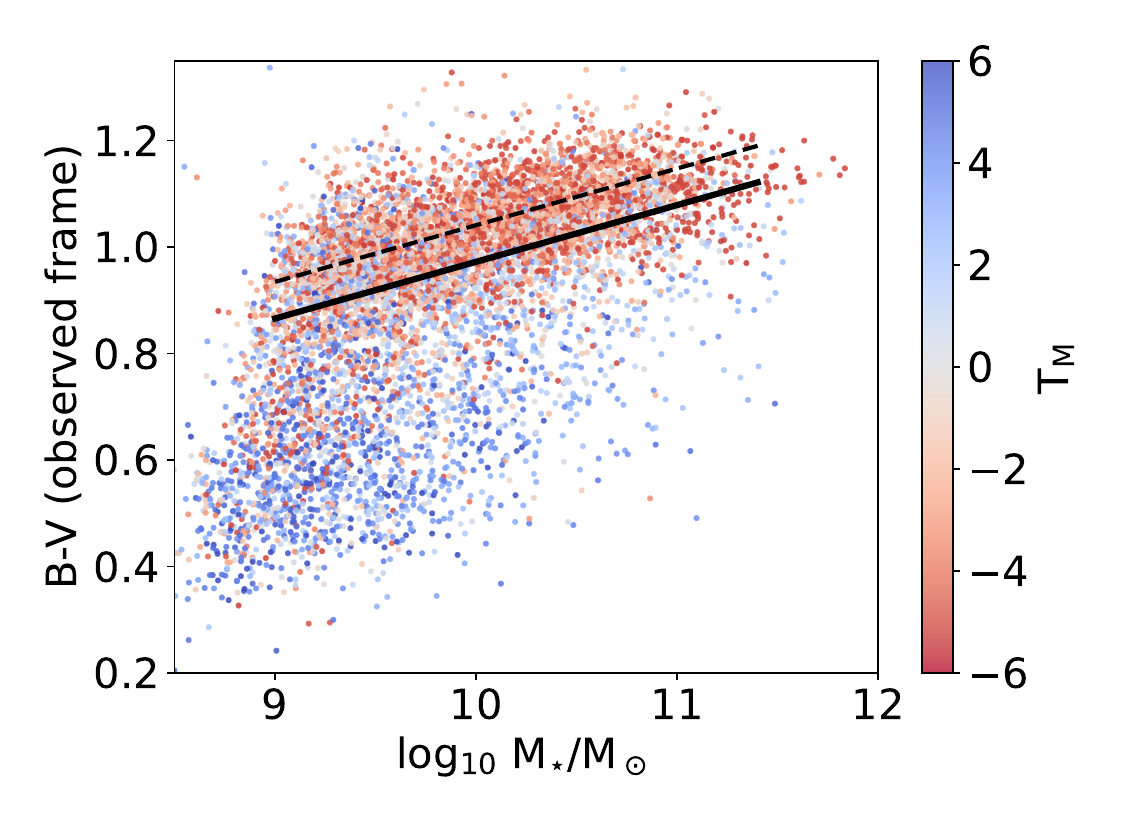}
\caption{Colour-mass plot for the OmegaWINGS cluster members, colour-coded by the MORPHOT type $T_{\rm M}$. The red sequence ridge and the blue/red separation are shown as dashed and solid black lines, respectively.}
\label{f:color_mass_diagram}
\end{center}
\end{figure}

We now discuss the morphological evolution of the GASP clean sample defined in Section \ref{sss:GASP}. By construction, given the selection criteria described in Section \ref{sss:GASP}, the great majority of galaxies in this sample are spirals, with the addition of a few early-type galaxies, mostly lenticulars.
We stress that, amongst the $3549$ synthetic galaxy images produced for this study, only $156$ ($\sim4\%$) were labelled as unclassified by our classification scheme and excluded from our analysis.

The left panel of Fig.\,\ref{f:morph_evo_all_stat} shows how the fraction of Spi, Len and Ell, determined as discussed in Section \ref{sss:classif_uncertainties}, varies with time for the three quenching scenarios considered.
The typical error estimates for these fractions is represented by the red error-bar ($\pm4\%$), determined from Fig.\,\ref{f:classification_error}.
All scenarios feature a decrease in the fraction of Spi with time, balanced by a corresponding increase in the fraction of Len (mostly) and Ell (to a lesser extent), until a plateau is reached where the fractions do not vary any longer.
By construction all scenarios begin with the same class distribution: $2\%$ Ell, $11\%$ Len, $87\%$ Spi.
We notice that, within the quoted uncertainty of $\sim4\%$, this is compatible with the MORPHOT classification ($2\%$, $7\%$, $91\%$), which is available for 54 of the 91 galaxies that build the GASP clean sample\footnote{MORPHOT types have not been computed for GASP galaxies in the PM2GC catalogue.}.
After the plateau is reached, all scenarios end with about similar distributions ($7\!-9\%$ Ell, $37\!-\!41\%$ Len, $50\!-\!56\%$ Spi).
Thus, about $40\%$ of the spirals are turned into early-type galaxies, mostly lenticulars.
Overall, the instantaneous quenching case retains a marginally higher fraction of spirals and produces less lenticulars, which is expected given that the outside-in scenarios promote concentration enhancements.
The most relevant difference between the three scenarios is the time at which the plateau is reached: as expected, this is shorter for the instantaneous quenching case ($\sim1.5\Gyr$), and becomes progressively longer for increasing stripping timescales ($\sim2\Gyr$ for $t_{\rm strip}\!=\!0.5\Gyr$, and $\sim3.5\Gyr$ for $t_{\rm strip}\!=\!1\Gyr$).

Even though the fraction of spirals remains large, the decline of their star formation inevitably causes an evolution in their colour.
Here we distinguish between `blue' and `red' galaxies, depending on their location on the $(B-V)$-$M_\star$ plane.
We separate the two populations in OmegaWINGS with the following approach, similar to that employed by \citet{Vulcani+22}. 
We first identify the red sequence \citep[e.g.][]{Baldry+04} by fitting with a straight line the upper `ridge' of the $(B-V)$-$M_\star$ plane, traced by the peaks of the galaxy number density in different $M_\star$ bins.
We then determine the characteristic `thickness` of the red sequence as the rms scatter around the line, computed using only galaxies above the ridge.
The blue/red separation is set by lowering the ridge line by this characteristic thickness, which gives
\begin{equation}\label{eq:red_vs_blue}
    (B-V)=0.106\times\log_{10}({M_\star/\msun})-0.092\,.
\end{equation}
Fig.\,\ref{f:color_mass_diagram} shows the $(B-V)$-$M_\star$ distribution for OmegaWINGS galaxies, together with the red sequence ridge and red/blue separation line.
The red sequence and its scatter do not vary much if we limited our analysis of the $(B-V)$-$M_\star$ plane to early type galaxies alone.
We note that, even though the galaxy sample used is similar, both the slope and the intercept of Equation (\ref{eq:red_vs_blue}) differ from those determined by \citet{Vulcani+22}, due to the fact that here we focus on the observed-frame colour.

In our evolutionary models, all galaxies will eventually quench and inevitably become red.
The right panel of Fig.\,\ref{f:morph_evo_all_stat} focuses on the spiral class alone, showing how the blue and red fractions evolve with time.
All scenarios show a similar trend, which is a systematic and almost complete transition from the blue cloud to the red sequence for all spirals in GASP.
The spiral population, initially made by blue systems only, becomes equally partitioned between red and blue systems after $\sim2\Gyr$.
As expected, the colour transition occurs faster for more efficient quenching scenarios, although the differences are not substantial: $\sim90\%$ of the spirals become red after $3\Gyr$ in the instantaneous quenching case, and after $4\Gyr$ in the two stripping cases.

Interestingly, the transformation in the colour of the spiral population proceeds well beyond the time at which the morphology distribution stabilises, especially for the most efficient quenching scenarios. 
In the instantaneous stripping case, after only $1\Gyr$, the morphology distribution has almost reached the plateau whereas the fraction of red spirals is still about zero. 
This indicates that there is a delay between the morphological and the colour transformation of quenched spirals, which is longer for more efficient quenching mechanisms.
The same result can be inferred by the fact that the color evolution of lenticulars (not shown here) is very similar to that of the spirals, indicating that, in our evolutionary models, morphology transformation typically occurs before galaxies move to the red sequence.
However, this result is heavily dependent on the adopted separation between blue and red galaxies: lowering the zero-point of the black solid line in Fig.\,\ref{f:color_mass_diagram} by $0.07$ dex (that is, assuming a thickness for the red sequence equal to twice the rms scatter measured) leads to a faster colour evolution, with red/blue equipartition reached after $1.2\Gyr$ and colour transformation fully completed after $2.5\Gyr$ for all scenarios.

Our results indicate that the simple ageing of the stellar population is a relevant factor in driving the morphological evolution of quenched cluster galaxies.
The quenching of a galaxy population originally dominated by blue spirals produces a mixed population made by red spirals ($50\!-\!55\%$) and lenticulars ($\sim40\%$) on timescales ranging between $1.5$ and $3.5\Gyr$, depending on the quenching scenario considered.
While a more thoughtful analysis is needed to model the transformation of a realistic cluster galaxy population, we remark that the morphology fractions in clusters are observed to evolve on similarly short timescales \citep[e.g.][]{Fasano+00, Donofrio+15}.

\subsection{The effect of stellar mass and environment}\label{ss:results_subGASP}
\begin{figure*}
\begin{center}
\includegraphics[width=\textwidth]{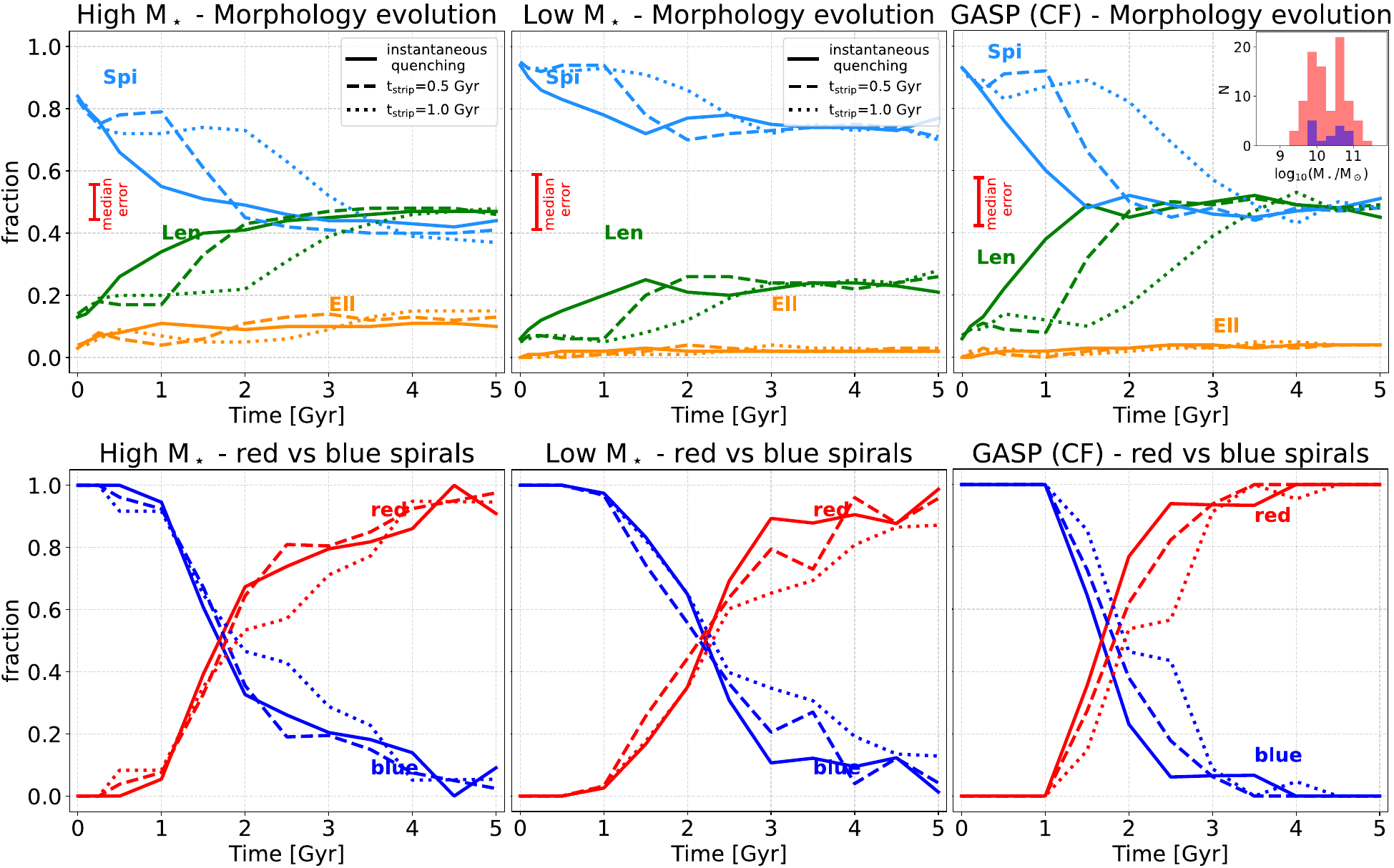}
\caption{As in Fig.\,\ref{f:morph_evo_all_stat}, but for GASP galaxies with $M_\star\!>\!10^{10.5}\msun$ (\emph{left column}), GASP galaxies with $M_\star\!<\!10^{10}\msun$ (\emph{central column}), and for the GASP control field (CF) sample (\emph{right column}). The inset in the top-right corner shows the $M_\star$ distribution of the GASP clean (red) and CF (blue) samples.}
\label{f:subsamples}
\end{center}
\end{figure*}
The GASP clean sample is composed by a combination of optically disturbed and undisturbed systems selected from different environments, and it is certainly not representative for the galaxy population that is newly accreted onto present-day clusters.
Although we postpone the modelling of realistic galaxy populations infalling onto clusters to a follow-up study, in this Section we discuss the role played by stellar mass and environment on the evolutionary results presented above.

Fig.\,\ref{f:subsamples} shows how the morphology and colour evolution varies between three different subsamples, namely a low-$M_\star$ sample made by systems with $M_\star\!<\!10^{10}\msun$, corresponding to the first tertile of the $M_\star$ distribution (left column), a high-$M_\star$ sample made by galaxies with $M_\star\!>\!10^{10.5}\msun$, corresponding to the last tertile of the $M_\star$ distribution (central column), and the control field (CF) sample of \citet{Vulcani+18a}, already introduced in Section \ref{sss:GASP} (right column).
The three samples show substantial differences in terms of morphology evolution.
The high-$M_\star$ population evolves more strongly than the low-$M_\star$ one, moving from an initial $4\%$ Ell, $14\%$ Len, $82\%$ Spi, to a final $13\%$ Ell, $46\%$ Len, $41\%$ Spi.
In comparison, variations in the low-$M_\star$ population are more modest (from $0\%$ Ell, $5\%$ Len, $95\%$ Spi to $3\%$ Ell, $26\%$ Len, $71\%$ Spi).
However, the timescales for reaching the morphology plateau are comparable between the two populations, and similar to those discussed for the full sample in Section \ref{ss:results_GASP}.
Differences in the colour evolution for spirals (and also for they other types, not shown here) are present but marginal, with the high-$M_\star$ population showing a slightly sharper transition from the blue cloud to the red sequence. 
Noticeably, the sample that features the strongest morphology transformation is the CF, which evolves from an initial $0\%$ Ell, $7\%$ Len, $93\%$ Spi to a final $3\%$ Ell, $49\%$ Len, $48\%$ Spi.
This effect is not driven by differences in stellar mass, given that the clean and CF samples have comparable $M_\star$ distribution, as shown in the inset in the top-right panel of Fig.\,\ref{f:subsamples}.
The CF also shows the sharpest transition in colour.

What drives the differences between the various subsamples?
To answer this question, we have inspected the evolution of the four morphological diagnostic parameters in the different samples, finding that $S$ plays an important role.
Low values of $S$ correspond to smoother galaxies and are typical of the lenticular and elliptical classes (Figs.\,\ref{f:OW_len} and \ref{f:OW_ell}). 
At the beginning of their evolution, all CF galaxies have $S\!<\!1.3$, whereas only $60\%$ of the clean GASP sample have $S\!<\!1.3$, and individual systems can have $S$ as high as $4$.
Clearly, the higher $S$ of GASP galaxies is due to a selection bias, given that the sample is mostly made by optically disturbed systems featuring tails and unilateral debris in their outskirts.
Recent star-formation bursts promoted by ram pressure \citep[e.g.][]{Vulcani+18a} can also contribute to higher $S$ values within the disc.
The evolution of $S$, instead, is strongly influenced by the initial galaxy structure, which in turns correlates with $M_\star$. 
The low-$M_\star$ sample is made by several irregular systems, whose substructures get amplified with time due to the lack of dynamical evolution, as already mentioned in Section \ref{ss:results_JO201} (see also Section \ref{ss:limitations}).
This effect is weaker in the high-$M_\star$ sample, which is dominated by grand-design spirals.
As a consequence, we find that $S$ tends to grow with time in low-mass galaxies, and to decrease in high-mass galaxies.
High-$M_\star$ systems also have tendency to reach higher $C$ values than low-$M_\star$ ones.

In summary, while it is certainly plausible that the morphological evolution of quenched galaxies depends on their stellar mass and environment, we believe that the differences between the various evolutionary trends presented in this Section might be affected by the (artificial) amplification of substructures, which is intrinsic to our method.
This limitation weakens the transition towards earlier galaxy types, and is more severe in galaxies featuring an irregular and clumpy morphology.

\section{Discussion}\label{s:discussion}
\subsection{Does the ageing of quenched galaxies lead to the observed cluster galaxy population?}
As anticipated in Section \ref{s:intro}, our final goal is to quantify the importance of the spectrophotometric evolution due to stellar ageing in producing the galaxy population that we observe in low-$z$ clusters.
For this purpose, the morphology evolution resulting from our analysis must be coupled with a cosmological framework that describes the growth of clusters with time, specifying the infall rate and the main properties (stellar mass, morphological type) of the newly accreted galaxies as a function of redshift.

While we postpone this investigation to a more detailed follow-up study, preliminary considerations can be drawn from the results of this paper.
For instance, ageing seems quite inefficient in producing ellipticals, while it is very efficient at transferring blue spirals to the red sequence, regardless of the quenching scenario or sample considered (Figs.\,\ref{f:morph_evo_all_stat} and \ref{f:subsamples}).
From MORPHOT, we know that the $M_{\star}\!>\!10^{10.5}\msun$ galaxy population in OmegaWINGS is made of $42\%$ Ell, $59\%$ Len, and $28\%$ Spi ($40\%$ on the red sequence, $60\%$ on the blue cloud).
Comparing these fractions with those shown in the left-hand panels of Fig.\,\ref{f:subsamples}, we conclude that the ageing of a star-forming galaxy population infalling onto clusters is likely to play a negligible role in building the substantial population of ellipticals observed in the densest environments on the Universe, even for the most extreme quenching scenarios.
Thus other processes such as mergers (prior to or following the cluster infall) must be invoked.
Also, not all observed spirals in clusters are red as we predict from the evolution of single, GASP-like quenched population, due to the infalling population of star-forming galaxies which accrete onto clusters at every epoch, possibly coupled with a broad distribution of quenching timescales.

\subsection{Limitations of our approach}\label{ss:limitations}
Perhaps the most limiting aspect of our analysis is its neglect for galaxy dynamics, in favour of a pure spectrophotometric approach.
We discuss the various implications of this assumption below, but stress from the beginning that dynamical effects tend to promote the transition towards earlier galaxy types. 
In a sense, our assumption is a conservative one: it leads us to underestimate the fraction of ellipticals and lenticulars formed by stellar ageing.

We have already introduced the problem of substructure amplifications (Section \ref{ss:results_JO201}), which is more severe in irregular/clumpy systems (Section \ref{ss:results_subGASP}).
To illustrate this problem more clearly, let us consider the case of a galaxy with a constant SFR, which is what we assume in our stripping scenarios in the regions within the stripping radius. 
In the presence of dynamical evolution, regions that are star forming at the observed epoch ($t\!=\!0$) will spread across the disc and slowly fade with time, while being replaced by new regions of star formation, mostly located on the galaxy's spiral arms. In our treatment, instead, mass-growth is limited by construction to regions that are star forming at $t\!=\!0$, which will inevitably enhance the intensity-contrast and promote the presence of substructures in the mock images (see the case of JO201 in Section \ref{ss:results_JO201}).
Thus our approach leads to overestimate the clumpiness parameter $S$, and consequently to underestimate the production of lenticulars and ellipticals, especially for longer continuations of the SFH.
This effect is further amplified by the fact that several GASP galaxies have $S$ values above the average (by selection), and that many low-$M_\star$ star-forming galaxies have an intrinsic irregular morphology (Section \ref{ss:results_subGASP}).
A robust quantification of this effect is difficult to obtain, but comparing Fig.\,\ref{f:morph_evo_all_stat} with the various panels of Fig.\,\ref{f:subsamples} can provide a sense of its magnitude.
In general, it is arguable that our approach is more suited to model the evolution of undisturbed star-forming galaxies with a regular structure, rather than for highly disturbed, clumpy systems.

Dynamical evolution is expected to have additional effects on the structure of a galaxy.
When a spiral galaxy subject to ram-pressure stripping is deprived of its dynamically coldest component (that is, its interstellar medium), resonances induced by spiral arms drive up the level of random motion in the stellar disc, which consequently becomes unable to support the spiral pattern itself \citep[e.g.][and references therein]{Sellwood+14}.
Resonances also amplify the stellar velocity dispersion \citep[but only in the direction parallel to the disc, see][]{Sellwood+13} without affecting the total angular momentum, leading to a smaller $v/\sigma$.
Thus internal dynamical evolution affects the stellar structure of the galaxy, pushing it towards the lenticular class.
The question is on which timescale this transformation occurs.
\citet{SellwoodCarlberg84} suggested that the spiral pattern would fade on a timescale of about $10$ disc rotations.
For a galaxy like the Milky Way this corresponds to $1\!-\!2\Gyr$, thus compatible with the timescales for morphological changes inferred in the current work.
In their numerical models, \citet{Fujii+11} found instead long-lived ($\sim10\Gyr$) spiral structures, even in the absence of dissipational effects of the interstellar medium.
These contradictory results do not allow us to draw firm conclusions, but we can safely state that resonances induced by spiral arms are not expected to affect the galaxy structure on timescales that are significantly shorter than those of the spectrophotometric evolution inferred here.

Even though galaxy mergers are disfavoured in dense environments, repeated close encounters of a cluster galaxy with other cluster members can promote bar formation, starburst episodes and angular momentum losses.
The cumulative effect of this galaxy `harassment' \citep[e.g.][]{Moore+96} is to transform an infalling low-mass spiral into a dwarf elliptical in a timescale of some Gyr. 
The efficiency of the harassment depends strongly on the mass, size and trajectory of the infalling system, being maximal for diffuse, low-mass galaxies on strongly radial orbits \citep[e.g.][]{Moore+99,Mastropietro+05,Bialas+15}.
While harassment can explain the observed abundance of early-type dwarfs in clusters \citep{Binggeli+87}, galaxies in the $M_\star$ range covered by GASP ($>10^{9.2}\msun$) should hardly be affected by this mechanism \citep[e.g.][]{Smith+10}.

Aside from the lack of dynamical evolution, our models make use of other simplifications.
One is the assumed continuation of the star formation history.
We have modelled the effect of ram pressure stripping as an idealised outside-in shrinking of the active star formation region: star formation is halted in regions outside of a circle whose radius exponentially decreases with time, while proceeding at a constant rate in the regions inside.
However, ram pressure exerted by the ICM has more complex effects on the interstellar medium and star formation properties of a cluster galaxy.
Both observations \citep{Merluzzi+13,Vulcani+18a, Vulcani+20} and simulations \citep{Tonnesen+09,Bekki+14} indicate that, prior to complete gas removal, ram pressure can enhance the star formation activity by compressing the gas and promoting the conversion of \hi\ into H$_2$ \citep{Moretti+20b,Moretti+20a}.
Also, ram pressure can induce radial flows that funnel gas towards the centre \citep{SchulzStruck01,RamosMartinez+18,Akerman+23}, especially in the presence of a bar \citep{Sanchez-Garcia+23}.
These flows can promote the growth of the supermassive black holes \citep{Poggianti+17b}, which would explain the excess of AGN incidence in cluster jellyfishes compared to undisturbed galaxies in the field \citep{Peluso+22}.
An increased star formation activity, especially if confined to the central galaxy regions, could promote the growth of the concentration index $C$, thus facilitating the transition towards earlier morphological types.
Also, AGN-driven feedback can suppress star formation in the central regions \citep[as it seems to be the case for JO201, see][]{George+19}, providing an additional quenching mechanisms that proceeds inside-out, thus in the opposite direction with respect to that explored in this study.
For simplicity we did not consider these additional effects in our models.
However, we stress that the instantaneous quenching scenario must be considered a limiting case, and the inclusion of an additional quenching mechanism such as that due to AGN feedback can only lead to an evolution closer to that predicted by such an extreme scenario.
Also, the mean star formation enhancement induced by ram pressure is expected to be a factor $1.4\!-\!1.7$ \citep{Vulcani+18a}, which is unlikely to affect our findings\footnote{although it can reach a factor of $\sim5$ for individual galaxies at the peak of their stripping process}.
Moreover, it is likely that GASP cluster galaxies that are are at a late stage of their stripping have already experienced all processes described above and started quenching, which justifies our modelling of their future SFH. 
Excluding the CF sample, about half of the remaining galaxies studied here are classified as in a strong stripping stage, or beyond (Jtype $\geq1$, Poggianti et al., in prep.).
Even though these considerations are only qualitative, they suggest that second-order effects produced by ram pressure, which are not accounted for in this study, are not expected to have a major impact on our results.

\subsection{How robust and universal is our classification strategy?}\label{ss:is_classification_robuts}
Our morphology classification method (Section \ref{ss:morphology}) deserves further discussion.
In an era where morphology classification relies on citizen science \citep[e.g.][]{Lintott+08,Schawinski+14} and machine learning algorithms \citep[e.g.][]{Sanchez+18}, we have preferred a simpler implementation based on a continuous function of four variables, calibrated on previously classified galaxies from the OmegaWINGS survey.

Our approach has two main advantages.
The first is that the use of a continuous function ensures that the probabilities defined in Eq.\,(\ref{eq:probability}) vary smoothly as a function of our parameters.
This is ideal for our analysis, which is built on the classification of objects whose main features evolve smoothly, and weakly, from one snapshot to another, as shown in in Fig.\,\ref{f:JO201_evo_maps}.
The second advantage is that we use a limited number of parameters, all with a clear physical significance and well documented connection to galaxy morphology.
The resulting 4D probability space can be visually inspected (Fig.\,\ref{f:OW_len}) in order to assess the importance of each parameter in setting the morphological class.
This is not always possible for more complex parameter spaces or, in general, for machine learning approaches.

The OmegaWINGS-based calibration ensures that the classification is adequate for galaxies in low-$z$ clusters, and is clearly the optimal choice for the GASP sample which, to a great extent, is covered by the OmegaWINGS survey.
However, caution must be used in producing the synthetic images: as the calibration is based on OmegaWINGS $B$- and $V$- band observations, we must produce mock images with compatible spatial resolution and signal-to-noise ratios.
This makes sure that the typical classification errors will be those reported in Fig.\,\ref{f:classification_error}, which are small enough to justify our classification strategy.
We stress that, amongst our diagnostic parameters, some are less robust than others against variations in signal-to-noise.
For instance, lower quality images may strongly affect the clumpiness parameter $S$ by making spiral structures and isolated star-forming clumps unrecognisable, leading to lower $S$ values (smoother galaxies) and, in turn, to an apparently higher fraction of lenticulars (Fig.\,\ref{f:OW_len}).
Image degradation naturally occurs during the evolution studied here, given that the ageing of stellar populations decreases the galaxy light-to-mass ratio, and, consequently, the signal-to-noise of synthetic images.

Considering the many caveats associated with $S$, one may ask whether adopting a simpler classification strategy based only on $M_\star$, $C$ and the $(B-V)$ could be more convenient.
We tested this by repeating our analysis excluding the $S$-related term in Eq.\,\ref{eq:metric}.
We found morphology evolution trends compatible with those shown in Fig.\,\ref{f:morph_evo_all_stat} within the quoted uncertainty, but the plateau is never fully reached and the transformation appear to proceed also at later times (but at a very slow pace), leading to a higher final fraction of early type galaxies.
The exclusion of $S$ in the metric has the advantage of virtually zeroing 
the fraction of unclassified images.
However, classification errors increase slightly (by a factor $\sim1.12$) with respect to those shown in Fig.\,\ref{f:classification_error}.
Also, visual inspection of the synthetic images shows various cases of red spirals that are incorrectly classified as lenticulars by the 3-parameter classification method, given that the presence of a spiral structure affects the disc smoothness and cannot be captured without accounting for $S$.
Therefore, the choice of including $S$ in the list of parameters that regulate our classification scheme is a conservative one, as it reduces the fraction of early-type galaxies formed by the evolution of late-type systems. 

A different question is whether our classification scheme is limited to nearby galaxy clusters and their infall regions (that is, the galaxy population traced by WINGS and OmegaWINGS), or is applicable also to fields and groups.
The existence of a morphology-density relation implies that environments of different densities show a diverse distribution of morphology classes \citep{Dressler+80, Vulcani+23}, but whether the class partitioning stays constant at given $M_\star$, $(B-V)$, $C$ and $S$ in differently environments remains to be investigated.
However, testing this goes beyond the purpose of this study.

\section{Summary and conclusions}\label{s:conclusions}
Spirals whose star formation has been quenched by environmental mechanisms evolve passively by rearranging both their structure, via gravitational processes, and their light, via the ageing of their stars.
In this work we have focused on the stellar ageing, and studied its effectiveness in driving a morphological transformation of a population initially made by late-type galaxies into earlier morphological types.

Our method, sketched in Fig.\,\ref{f:flow_chart}, makes use of two ingredients.
The first is the modelling of the spectrophotometric evolution of galaxies with known spatially resolved SFH. 
For this purpose we used the galaxy sample from the GASP program \citep{Poggianti+17} (91 systems after the removal of merging, quenched and interacting objects), for which resolved SFH, extinction and metallicity maps have been derived in previous works by applying the \textsc{sinopsis} spectral modelling software on the GASP MUSE data \citep{Fritz+11,Fritz+17}.
We have produced synthetic spectroscopic datacubes and images for the evolved GASP galaxies assuming different continuations of their future SFH, instantaneous or progressive outside-in quenching.
The second ingredient is a tool for determining the morphological class of our synthetic galaxies. 
This tool gives the probability that a galaxy belongs to a given morphological class (elliptical, lenticular, spiral) as a function of four parameters (stellar mass $M_\star$, observed-frame colour $B-V$, concentration $C$, clumpiness $S$), and is calibrated on nearby cluster galaxies from the OmegaWINGS survey \citep{Gullieuszik+15, Moretti+17} with highly refined morphological type measurements from the MORPHOT software \citep{Fasano+12, Vulcani+23}.
By combining these ingredients, we followed the morphological evolution of the GASP population as a whole, and of a number of its subsamples, for $5\Gyr$.
Our results are the following:
\begin{itemize}
    \item individual star-forming galaxies such as JO201 show a steady growth of $(B-V)$ and $C$ with time caused by the ageing of their young stellar population. The ageing produces a redder and fainter disc, leading to a larger bulge-to-disc ratio. From a blue grand design spiral, JO201 turns into a lenticular after $0.9\!-\!1.8\Gyr$, depending on the scenario considered (Figs.\,\ref{f:JO201_evo_maps} and \ref{f:JO201_evo_diagnostic}).
    
    \item As a whole, the GASP population rearranges its morphological distribution from an initial ($2\%$ Ell, $11\%$ Len, $87\%$ Spi) to a final ($7\!-9\%$ Ell, $37\!-\!41\%$ Len, $50\!-\!56\%$ Spi).
    The transformation is completed after $1.5\!-\!3.5\Gyr$, occurring faster for more efficient quenching scenarios (left panel of Fig.\,\ref{f:morph_evo_all_stat}).
    
    \item During their evolution, GASP spirals do not only decrease in number, but also systematically move from the blue cloud to the red sequence. The spiral population, originally made by blue systems, becomes equally partitioned between red and blue objects after $\sim2\Gyr$, and fully dominated by red-sequence galaxies by the end of the $5\Gyr$ period. This colour evolution depends only weakly on the quenching scenario assumed (right panel of Fig.\,\ref{f:morph_evo_all_stat}), although it is sensitive to the red/blue separation adopted (Fig.\,\ref{f:color_mass_diagram}).


    \item The morphological transformation is stronger for the high-mass ($M_\star>10^{10.5}\msun$) population and for the optically undisturbed galaxies of the control field sample (Section \ref{ss:results_subGASP}). We interpret this result as a side-effect of the lack of dynamical evolution in our models, which leads to the growth of galaxy clumpiness $S$ and slows down the transition to earlier morphology types in galaxies that feature a highly clumpy or irregular optical morphology.    
\end{itemize}

As dynamical processes tend to promote the
transition towards earlier galaxy types, their neglect implies that our results are particularly conservative, as the fraction of
ellipticals and lenticulars formed by stellar ageing is likely underestimated.
Also, we believe that the artificial amplification of substructures is the main side-effect of the lack of dynamical evolution in our models: galaxy harassment is expected to have a very limited impact on galaxies as massive as those in the GASP sample \citep{Smith+10}, while secular dynamical processes that lead to the fading of the spiral structure and to a diminished rotational support in gas-less spirals act on timescales of $1\!-\!10\Gyr$ \citep{SellwoodCarlberg84,Fujii+11}, similar to or larger than those inferred here for the spectrophotometric evolution.

This work is preparatory to a more complex study that aims to determine to which extent the morphological evolution of spirals discussed here is responsible for the morphology distribution that we observe today in nearby clusters.
The answering of this question requires combining the results of this study with an infall model that characterises the population of galaxies accreted by clusters as a function of time, and is the subject of our next project.

\section*{Acknowledgements}
 Based on observations collected at the European Organization for Astronomical Research in the Southern Hemisphere under ESO program 196.B-0578.
This project has received funding from the European Research Council (ERC) under the European Union’s Horizon 2020 research and innovation program (grant agreement No.\,833824).

\section*{Data Availability}
The data underlying this article will be shared on reasonable request to the corresponding author.


\bibliographystyle{mnras}
\bibliography{main} 




\appendix
\section{Additional properties of OmegaWINGS galaxies}\label{app:pspi_pell}
\begin{figure*}
\begin{center}
\includegraphics[width=0.9\textwidth]{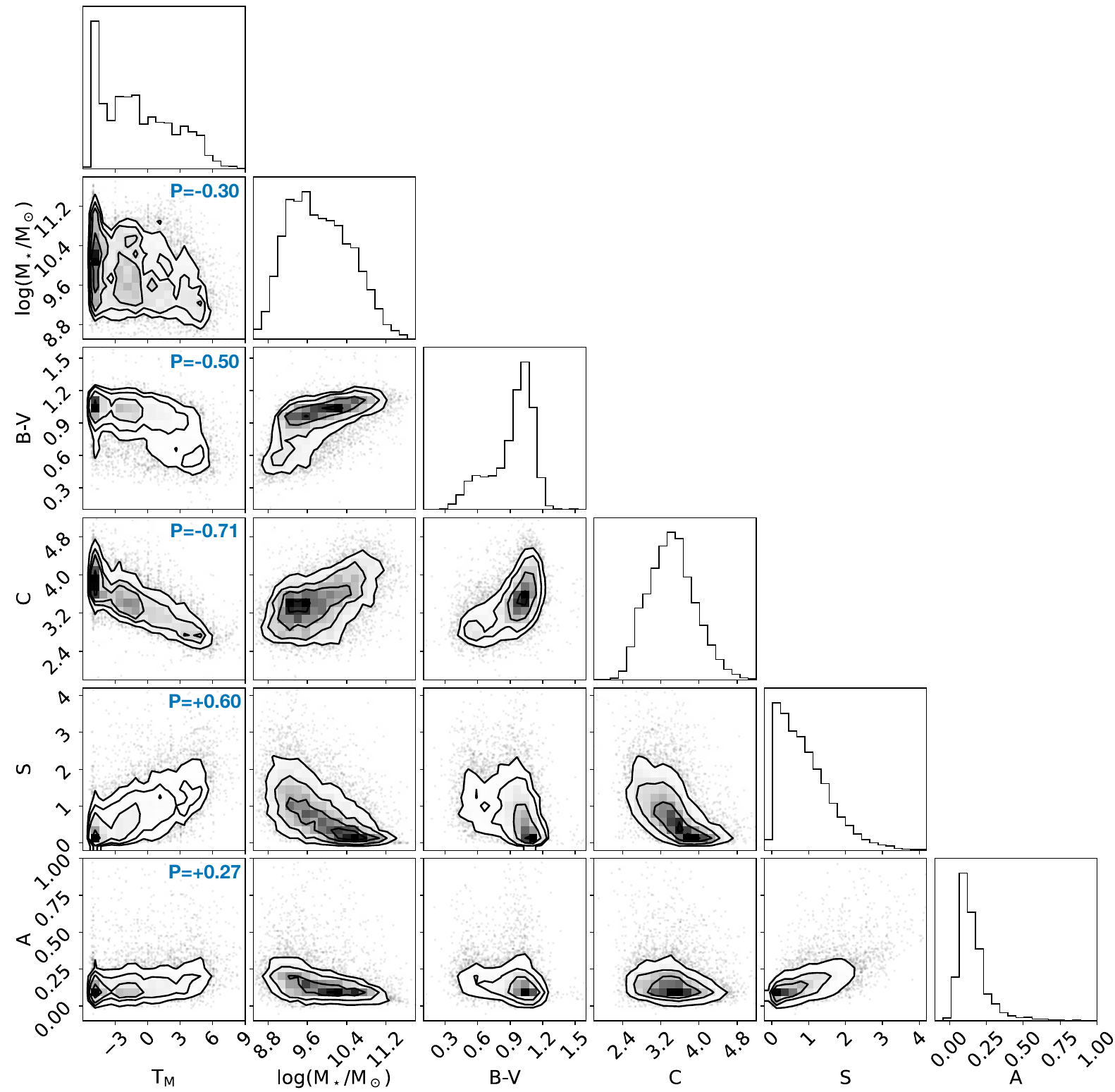}
\caption{Corner-plots showing the correlation between the various parameters (shaded regions, with contours at arbitrary iso-density levels), along with their overall distribution (histograms on top), for OmegaWINGS galaxies. The first column focuses on the correlations with $T_{\rm M}$, with the Pearson correlation coefficient reported in the top-right corner of each panel.}
\label{f:corner}
\end{center}
\end{figure*}

Fig.\,\ref{f:corner} shows the correlation between different parameters (MORPHOT type $T_{\rm M}$, stellar mass $M_\star$, observed-frame $B-V$ colour, concentration $C$, clumpiness $S$, asymmetry $A$) for galaxies in OmegaWINGS.
Morphology is most strongly correlated with $C$, as indicated by the Pearson coefficient (reported on the top-right corner in each panel of the first column), while it is poorly correlated with $A$.
The correlation between $T_{\rm M}$ and $M_\star$ is also relatively weak, but morphology types are well separated in the colour-mass plane shown in Fig.\,\ref{f:color_mass_diagram}, which supports our choice of $M_\star$ as an additional parameter for our classification scheme.

\begin{figure*}
\begin{center}
\includegraphics[width=\textwidth]{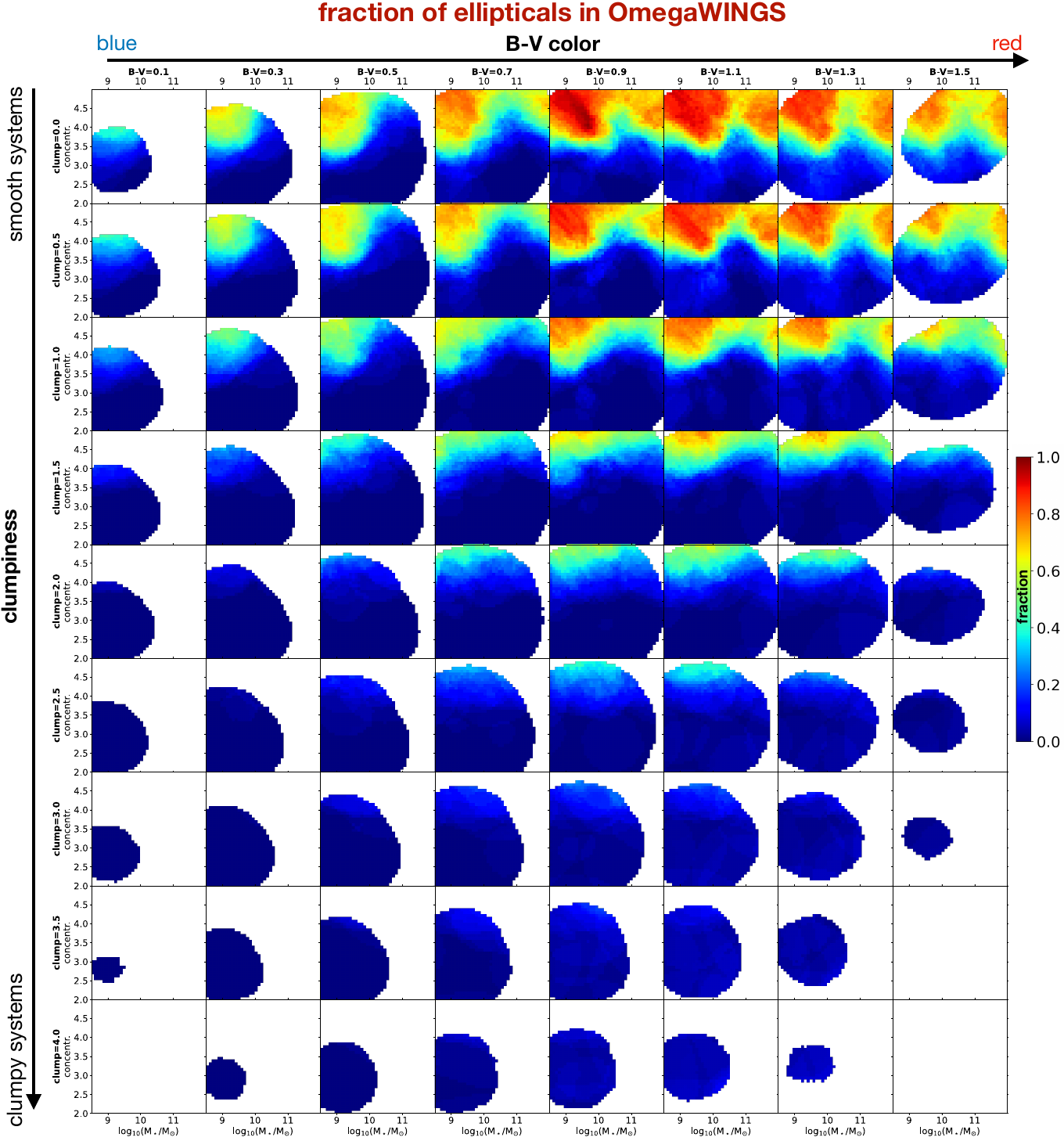}
\caption{As in Fig.\,\ref{f:OW_len}, but for elliptical galaxies, defined as systems with $-5.5\!<\!T_{\rm M}\!<\!-4.25$.}
\label{f:OW_ell}
\end{center}
\end{figure*}

\begin{figure*}
\begin{center}
\includegraphics[width=\textwidth]{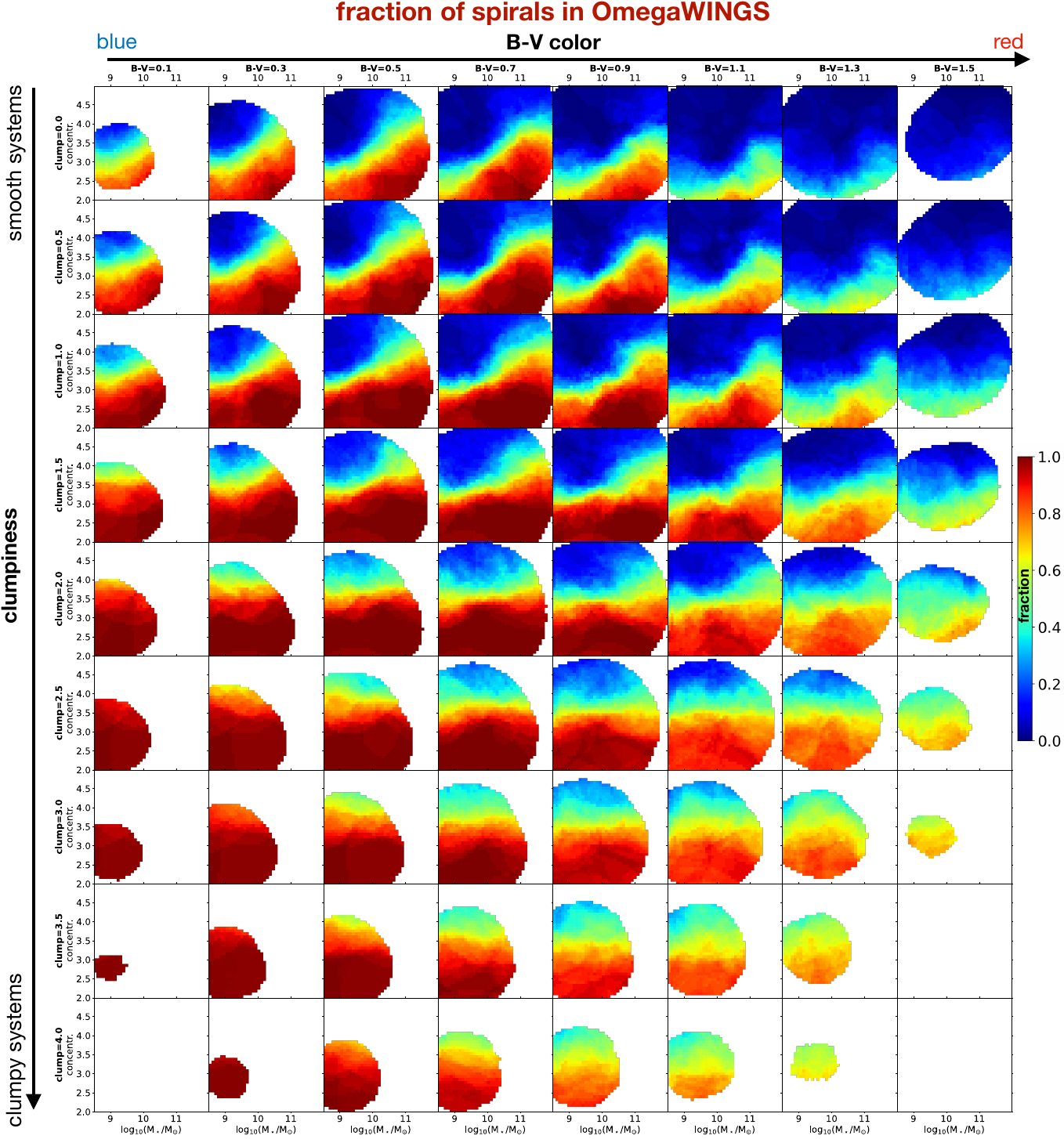}
\caption{As in Fig.\,\ref{f:OW_len}, but for spiral galaxies, defined as systems with $T_{\rm M}\!>\!0$.}
\label{f:OW_spi}
\end{center}
\end{figure*}

Figs.\,\ref{f:OW_ell} and \ref{f:OW_spi} are analogous to \ref{f:OW_len} and show, respectively, the $p_{\rm Ell}(\vec{x})$ and the $p_{\rm Spi}(\vec{x})$ determined for OmegaWINGS galaxies as described in Section \ref{sss:classif_method}. 

\bsp	
\label{lastpage}
\end{document}